\documentclass[aps,prd,amssymb,amsmath,amsfonts,superscriptaddress,nofootinbib,eqsecnum,reprint,showpacs,longbibliography]{revtex4-2}
\usepackage{graphicx}
\usepackage{bm}
\usepackage{comment}
\usepackage{color}
\usepackage{mathtools}
\usepackage{commath}
\usepackage{dcolumn}% Align table columns on decimal point
\usepackage{xcolor}
\usepackage{orcidlink}
\usepackage{hyperref}
\usepackage{cleveref}
\usepackage{orcidlink}
\usepackage{tensor}
\usepackage{booktabs,array,longtable,multirow,tabularx,hhline}

\begin{document}
\title{Dipolar boson stars as static two-body systems: an effective-one-body description}
\author{Thibault Damour}%
 \email{damour@ihes.fr}
    \affiliation{%
Institut des Hautes Etudes Scientifiques, 
91440 Bures-sur-Yvette, France
}%
\author{Romain Gervalle}%
 \email{romaingervalle@ua.pt}
    \affiliation{%
    Departamento de Matemática da Universidade de Aveiro and Center for Research and Development in Mathematica and Applications -- CIDMA, Campus de Santiago, 3810-183 Aveiro, Portugal
}%
    \affiliation{%
    Département de Physique de l'Université de Tours, Parc de Grandmont, 37200 Tours, France
}%
\author{Tamanna Jain}%
 \email{tj317@cam.ac.uk}
\affiliation{%
LPENS, Département de physique, Ecole normale supérieure, Université PSL, Sorbonne Université, Université Paris Cité, CNRS, 75005 Paris
}%
    \affiliation{%
 Department of Applied Mathematics and Theoretical Physics,University of Cambridge, Wilberforce Road CB3 0WA Cambridge, United Kingdom
}%
\date{\today}
\begin{abstract}
Dipolar boson stars consist of two scalar-field lumps held in static equilibrium by the balance between gravitational attraction and scalar repulsion. {The repulsion stems from a $\pi$ phase difference between the harmonically varying complex scalar field sourcing each lump.} We construct sequences of these configurations numerically and describe them analytically as bound states of two boson stars within the effective-one-body (EOB) formalism. We {map} the EOB potential previously derived for the scattering of two boson stars onto {an energy-dependent Schwarzschild-like radial potential.} %, better suited to bound configurations, by a simple energy-dependent radial-coordinate transformation. 
{We find that the EOB-based analytical prediction for the binding energy of the system agrees remarkably well with the numerical values. The analytical prediction for the critical gravitational coupling beyond which no static equilibrium exists is also consistent with the numerical results.}
%We find that the effective energy ($\gamma$), which is related to the binding energy of the system, predicted by the analytical EOB model agrees well with the numerical values. The model also predicts a critical gravitational coupling beyond which no static equilibrium exists, in agreement with the numerical results. 
\end{abstract}

\maketitle

\section{Introduction}

A century after its formulation, General Relativity (GR) continues to withstand every experimental test to which it has been subjected, particularly in the strong-field, dynamical regime probed by gravitational-wave astronomy. Since the first detection of a binary black hole merger in 2015~\cite{Abbott:2016blz}, the LIGO-Virgo-KAGRA network has accumulated several hundred confident gravitational-wave events, most recently compiled in the fifth Gravitational-Wave Transient Catalog (GWTC-5.0)~\cite{LIGOScientific:2026}, bringing the total to 390 detections.  

{Most gravitational-wave signals detected so far are consistent with the coalescence of binary black holes. In particular, the loudest signal observed to date, GW250114, enabled for the first time a multi-mode measurement of the ringdown of the remnant object, thereby providing a very stringent confirmation of the GR-predicted structure of black holes~\cite{LIGOScientific:2025rid}. The other events are consistent with the GR-predictions for the merger of  binary neutron stars or neutron star-black hole systems.} Nevertheless, the possibility that some of these signals instead originate from the coalescence of horizonless exotic compact objects remains open~\cite{Cardoso:2019rvt,Maggio2021}.  Scalar boson stars are among the most tractable examples of such objects, as they arise as regular, horizonless, self-gravitating solutions of the Einstein-Klein-Gordon equations~\cite{Kaup:1968zz,Ruffini:1969qy} (see Refs.~\cite{Liebling:2012fv,Schunck:2003kk} for reviews). Recent studies point to a significant degeneracy between signals from boson-star and black-hole binaries within the sensitivity of current detectors~\cite{Evstafyeva:2024qvp,Evstafyeva2026,Vangstein:2026osl} (see also Refs.~\cite{Bustillo:2020syj,CalderonBustillo:2022cja}, which interpret several gravitational-wave events as vector-boson-star mergers). 
Boson stars have also attracted increasing attention in other contexts, such as dark matter searches~\cite{Liebling:2012fv,Kim:2024wku}, structure formation~\cite{Vachaspati:1991tt,Inman:2019wvr}, and black hole mimickers~\cite{Guzman:2009zz,Cunha:2022gde,Siemonsen:2024snb,Marks2025}.

{The simplest boson stars are static, spherically symmetric, self-gravitating lumps of scalar-field energy sourced by a %{\color{violet}self-interacting} \rg{The simplest BS are not necessarily "self-interacting". The scalar field can be free of self-interactions.} 
harmonically varying complex scalar field $\Phi(r,t)=\phi(r) e^{- i \omega t}$.} They possess an established formation mechanism~\cite{Seidel:1993zk,Kasuya2000} and satisfy the criteria for dynamical stability over part of their domain of existence~\cite{Gleiser:1988rq,Gleiser:1988ih,Lee1989,Kusmartsev1991,Kusmartsev1992,Tamaki2011}. 
{Besides spherically symmetric boson stars, there also exist angular-momentum-carrying (rotating) axially symmetric boson stars \cite{Schunck:1996he,Yoshida:1997qf,Kleihaus:2005me} and, more surprisingly, \textit{dipolar boson stars}, i.e. axisymmetric, static systems consisting of two separate lumps of scalar-field energy density held in static equilibrium.}
 Dipolar boson stars were first described in the pioneering work of Yoshida and Eriguchi~\cite{Yoshida:1997nd} (and even earlier, in the Newtonian approximation, by Schupp and van der Bij~\cite{Schupp:1995dy}). They can be interpreted as two ``monopolar" boson stars in equilibrium, with their gravitational attraction balanced by a scalar repulsion arising from a relative phase difference of $\pi$~\cite{Palenzuela:2006wp,Cunha:2022tvk} between the scalar field above and below the equatorial plane separating the two lumps. Remarkably, when self-interactions are included in the scalar field potential, the dipolar boson stars exhibit similar stability property as the spherically symmetric stars, with in particular, a part of the parameter space where they can be dynamically stable~\cite{Ildefonso2023}.

The construction of dipolar boson stars is relatively straightforward from the numerical point of view: the relative phase difference of $\pi$ is imposed by requiring the {real (time-independent) scalar-field amplitude $\phi= e^{i \omega t} \Phi$} to be antisymmetric with respect to the equatorial plane. The solutions can then be obtained using standard numerical techniques for elliptic field equations. However, a consistent analytical description of the interaction between the two scalar lumps is still lacking in the literature. Reference~\cite{Cunha:2022tvk} considered a {heuristic} analytical model, combining a Newtonian potential for the gravitational attraction with a Yukawa term accounting for the scalar repulsion. {However, neither the range nor the strength of the scalar repulsion needed to balance the gravitational attraction were derived from first principles in Ref.~\cite{Cunha:2022tvk}; they were instead fitted to numerical data. In this work, we instead propose a more elaborate analytical description of the competing interactions using two first-principle ingredients: (i) the notion
of scalar charge $q_s$ (distinct from the Noether charge $Q$) sourcing the scalar field; and (ii)
the effective-one-body (EOB) formalism~\cite{Buonanno:1998gg, Buonanno:2000ef}.} A similar approach has  recently been applied to the gravitational scattering of boson stars~\cite{Damour:2025oys}, showing good agreement with numerical relativity simulations and paving the way toward analytic waveform templates of exotic compact object binaries. Since dipolar boson stars are static, the present work provides, for the first time, an application of the EOB formalism to describe a \textit{static} two-body interaction.

The rest of the paper is organized as follows. In Sec.~\ref{sec:model}, we introduce the (solitonic) boson star model and its global charges. In Sec.~\ref{sec:numerical}, we present our numerical approach to solving the field equations, together with the appropriate ans\"atze and boundary conditions; we also present in this section the numerical solutions describing isolated and dipolar boson stars, along with their physical observables. The analytical model describing the static two-body interaction, based on the EOB approach, is presented in Sec.~\ref{sec:eob_model}, while the comparison with the numerical results is given in Sec.~\ref{sec:num_vs_analytic}. Finally, our conclusions are presented in Sec.~\ref{sec:conclusions}.

Throughout the paper, we use the metric signature $(-,+,+,+)$ and units in which $c=\hbar=1$. {We do not set $G$ to 1, as we will vary %the relative strength of gravitational and scalar interactions by varying the dimensionless parameter $\hat G = G \Phi_0$, see below.
the strength of gravity relative to the scalar self-interactions through the dimensionless parameter $\hat G = G \Phi_0$, see below.}

\section{The model}
\label{sec:model}

We consider the theory of a complex scalar field $\Phi$ minimally coupled to Einstein's gravity. It is described by the following action
\begin{equation}
    \label{eq:action}
    S=\int{d^4 x\,\sqrt{-g}}\left\{\frac{R}{16\pi G}-g^{\alpha\beta}\partial_\alpha\Phi^\ast\partial_\beta\Phi-U(|\Phi|)\right\},
\end{equation}
where $R$ is the Ricci scalar of the spacetime metric $g_{\alpha\beta}$, $G$ is Newton's constant, and $U(|\Phi|)$ is the scalar field potential. In this work, we employ the so-called solitonic potential \cite{Friedberg1987}
\begin{equation}
    \label{eq:potential}
    U(|\Phi|)=\mu^2|\Phi|^2\left(1-\frac{|\Phi|^2}{\Phi_0^2}\right)^2,
\end{equation}
which  {adds to the  mass term $\mu^2|\Phi|^2$}  an attractive quartic and a repulsive sextic self-interaction. Here $\mu$ is the scalar field mass, and $\Phi_0$ is the location of the nontrivial zero of the potential characterizing the self-interactions.

The corresponding Einstein-Klein-Gordon equations are obtained by varying the action with respect to the metric and scalar field
\begin{align}
\label{eq:ein}
    R_{\alpha\beta}-\frac{1}{2}g_{\alpha\beta}R&=8\pi G\,T_{\alpha\beta}\,,\\
\label{eq:KG}
    \left(\nabla_\alpha\nabla^\alpha-\frac{dU}{d|\Phi|^2}\right)\Phi&=0,
\end{align}
where $R_{\alpha\beta}$ is the Ricci tensor and $T_{\alpha\beta}$ is the stress-energy tensor of the scalar field which reads
\begin{align}
    \nonumber
    T_{\alpha\beta}=&\,\,\partial_\alpha\Phi^\ast\partial_\beta\Phi+\partial_\beta\Phi^\ast\partial_\alpha\Phi \\
    &-g_{\alpha\beta}\left[\partial_\lambda\Phi^\ast\partial^\lambda\Phi+U(|\Phi|)\right].
\label{eq:stress_energy}
\end{align}

The model \eqref{eq:action} is invariant under the global $\text{U}(1)$ transformation $\Phi\to e^{i\lambda}\Phi$, with a constant $\lambda$. This symmetry is associated with a conserved current, $j^\alpha=-i\left(\Phi\partial^\alpha\Phi^\ast-\Phi^\ast\partial^\alpha\Phi\right)$ with $\nabla_\alpha j^\alpha=0$. Integration over a spacelike hypersurface $\Sigma$ of its timelike component gives the conserved Noether charge
\begin{equation}
\label{eq:Q_def}
    Q=\int_\Sigma{j_\alpha n^\alpha dV},
\end{equation}
where $n^\alpha$ is the future-directed unit normal vector to $\Sigma$. At the microscopic level, this quantity counts the number of scalar particles in the field configuration.

We are interested in static geometrical configurations where the metric admits a timelike Killing vector $\xi=\partial_t$ (in adapted coordinates). We consider boson star solutions that are spatially localized and have finite mass, hence, spacetime is asymptotically flat. {The total mass is given by the Tolman-Komar volume integral~\cite{Wald1984},}
\begin{align}
\nonumber
    M=&\,
    \frac{1}{4\pi G}\int_\Sigma{R_{\alpha\beta}n^\alpha\xi^\beta\,dV}\\
    =&\,
    2\int_\Sigma{\left(T_{\alpha\beta}-\frac{1}{2}g_{\alpha\beta}T\right)\,n^\alpha\xi^\beta\,dV}.
\end{align}

A boson star configuration is parametrized by \textit{two dimensionless parameters} which can be taken to be,
\begin{equation}
    \hat{Q} \equiv\frac{Q}{\mu^2\Phi_0^2},\qquad\hat{G} \equiv G\,\Phi_0^2.
\end{equation}
Numerically, we work with units where $\mu=1$ and $\Phi_0=1$, so that $\hat{Q}=Q$ and $\hat{G}=G$. For simplicity, we shall henceforth often drop the hat notation.

\section{Numerical construction}
\label{sec:numerical}

In this section, we describe the numerical construction of the boson star solutions.
We employ the following metric ansatz describing a static and axially-symmetric spacetime
\begin{equation}
\label{eq:metric_ansatz}
ds^2=-e^{2F_0}dt^2+e^{2F_1}(dr^2+r^2 d\theta^2)+e^{2F_2}r^2\sin^2\theta\,d\varphi^2.
\end{equation}
The three functions $F_0$, $F_1$, and $F_2$ depend only on the quasi-isotropic radial coordinate $r$ and the polar angle~$\theta$. For the complex scalar field, we adopt the harmonic ansatz
\begin{equation}
\label{eq:scalar_ansatz}
\Phi=\phi(r,\theta)e^{-i\omega t},
\end{equation}
where $\phi$ is the (real) scalar-field amplitude and $\omega$ is the scalar field internal frequency. Although the scalar field itself depends explicitly on time, this harmonic dependence leaves the stress-energy tensor in Eq.~\eqref{eq:stress_energy} time-independent. Such harmonic time dependence is a characteristic feature of boson stars and allows non-trivial, localized configurations to exist by {circumventing Derrick's obstruction ~\cite{Derrick1964,Carloni:2019cyo} to static solitons.}

Substituting the line element \eqref{eq:metric_ansatz} and scalar field \eqref{eq:scalar_ansatz} into the Einstein--Klein--Gordon equations \eqref{eq:ein} and \eqref{eq:KG} yields a coupled system of four partial differential equations (PDEs) for the {bivariate} functions $\phi(r,\theta)$, $F_0(r,\theta)$, $F_1(r,\theta)$, and $F_2(r,\theta)$. This system is elliptic, and is solved numerically as a boundary-value problem, with appropriate boundary conditions imposed to ensure asymptotic flatness and regularity.

We are interested in two classes of solutions:
\begin{itemize}
    %\item \underline{\textit{Isolated boson stars:}} these are spherically symmetric scalar-field lumps and correspond to the ground state of static boson stars.
    \item \underline{\textit{Isolated boson stars:}} these are spherically symmetric self-gravitating scalar-field lumps. We restrict our attention to the ground state, for which the scalar field has no radial nodes. They can be obtained by setting $F_1=F_2$ in Eq.~\eqref{eq:metric_ansatz}, and requiring $\phi$, $F_0$, and $F_1$ to depend only on the radial coordinate $r$ (which then becomes exactly isotropic). The appropriate boundary conditions in this case are
    \begin{align*}
        \partial_r F_{0,1}\rvert_{r=0}=\partial_r\phi\rvert_{r=0}&=0,\\
        F_{0,1}\rvert_{r\to\infty}=\phi\rvert_{r\to\infty}&=0.
    \end{align*}
    \item \underline{\textit{Dipolar boson stars:}} these consist of two identical lumps of scalar-field energy density separated along a symmetry axis. The two lumps are held in static equilibrium by a balance between (long-range) gravitational attraction and (short-range) scalar repulsion~\cite{Yoshida:1997nd,Cunha:2022tvk}. They are obtained by requiring the scalar-field amplitude $\phi$ to be antisymmetric with respect to the equatorial plane, $\phi(r,\theta)=-\phi(r,\pi-\theta)$. Restricting the polar variable to $\theta\in[0,\pi/2]$ owing to this equatorial symmetry, the boundary conditions read
    \begin{align*}
        \partial_r F_{0,1,2}\rvert_{r=0}=\phi\rvert_{r=0}&=0,\\
        F_{0,1,2}\rvert_{r\to\infty}=\phi\rvert_{r\to\infty}&=0,\\
        \partial_\theta F_{0,1,2}\rvert_{\theta=0}=\partial_\theta\phi\rvert_{\theta=0}&=0,\\
        \partial_\theta F_{0,1,2}\rvert_{\theta=\pi/2}=\phi\rvert_{\theta=\pi/2}&=0.
    \end{align*}
\end{itemize}

In our numerical approach, we compactify the semi-infinite interval $r\in[0,\infty)$ using a new radial coordinate $x=r/(r+1)$, which allows us to integrate the field equations up to infinity without relying on a cut-off radius. For isolated boson stars, the computational domain then reduces to the finite segment $x\in[0,1]$, while for dipolar boson stars the field equations are solved on the rectangle $(x,\theta)\in[0,1]\times[0,\pi/2]$. 

We solve the field equations with the finite-element solver \texttt{FreeFem}~\cite{MR3043640}. This solver operates on the weak form of the PDEs, obtained by recasting them as integral equations. The unknown functions are expanded over elementary basis functions associated with a chosen discretization of the computational domain. The resulting discrete problem is a system of nonlinear algebraic equations, which we solve using a Newton-Raphson algorithm. We estimate the numerical accuracy of the solutions using a relativistic virial identity (see Eq.~(7.39) of Ref.~\cite{Gervalle2025}). The solutions constructed in this work typically reach an accuracy of $\sim10^{-14}$ for isolated boson stars and $\sim10^{-7}$ for dipolar boson stars. Further details on the application of this solver to strong-gravity problems can be found in Ref.~\cite{BSFactory}.

In the code, numerical solutions are specified by two input parameters: the gravitational coupling $G$ {(alias $\hat G = G \Phi_0$)} and the scalar-field frequency $\omega$ {(alias $\hat \omega =\omega/\mu$)}. The Noether charge $Q$ {(alias $\hat Q = Q/(\mu^2 \Phi_0^2)$)} is therefore not an input parameter but rather an output. For the ansatz \eqref{eq:metric_ansatz}-\eqref{eq:scalar_ansatz}, $Q$ is  obtained by evaluating the on-shell integral
\begin{equation}
    Q=4\pi\omega\int_0^\infty dr\int_0^{\pi}d\theta\,r^2\sin\theta\,e^{-F_0+2F_1+F_2}\,\phi^2.
\end{equation}
{Nevertheless, we are able to construct solutions for a given $Q$ (instead of a given $\omega$) %iteratively, using a script based on the quasi-Newton method, which adjusts $\omega$ until the desired value of $Q$ is reached.
by iteratively adjusting $\omega$ with a quasi-Newton method until the desired value of $Q$ is reached.} %This script requires running the code several times to obtain a single solution with the prescribed value of $Q$.

\begin{figure}
    \centering
    \includegraphics[width=\linewidth]{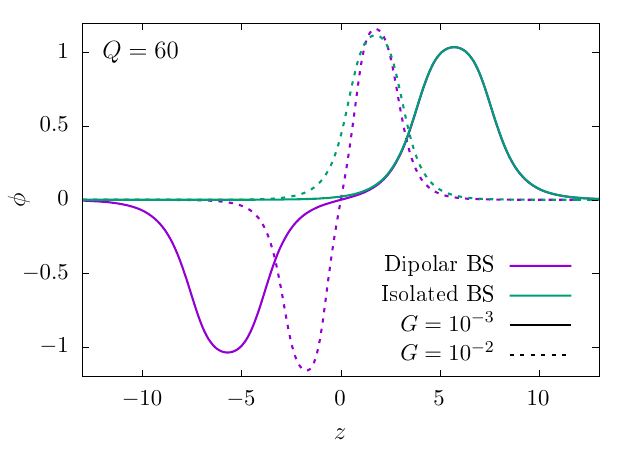}
    \caption{\small Comparison (in units $\mu=1=\Phi_0$) of the scalar profiles along the $z$-axis for an isolated boson star with Noether charge $Q=60$ (green curves) and a dipolar boson star with total Noether charge $Q_\text{dipole}=120$ (purple curves). Solid and dashed lines correspond to $G=10^{-3}$ (weak gravity) and $G=10^{-2}$ (strong gravity), respectively. The isolated boson star profiles are shifted along the $z$-axis so that their maximum coincides with that of the corresponding dipolar configuration (on the $z>0$ axis).}
    \label{fig:dbs_vs_ibs}
\end{figure}

\begin{figure*}[t]
    \centering
    \includegraphics[width=0.49\textwidth]{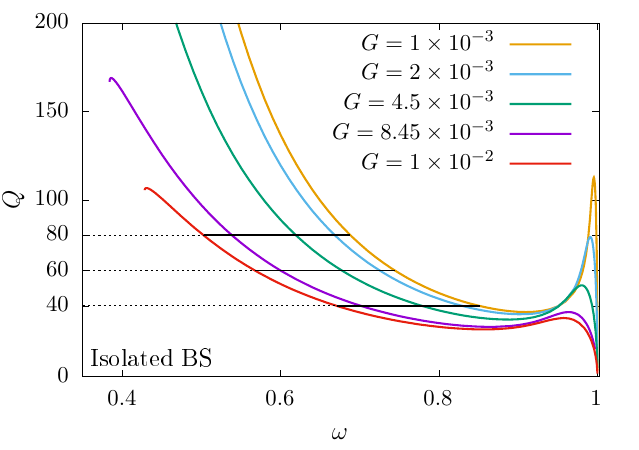}\hfill
    \includegraphics[width=0.49\textwidth]{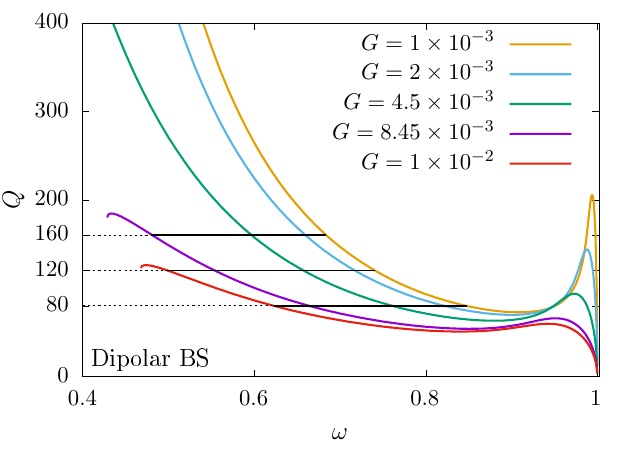}
    \caption{\small Noether charge $Q$ (in units of $(\mu\Phi_0)^2$) %$Q (= Q/(\mu^2 \Phi_0^2))$ 
    against the scalar-field frequency $\omega$ (in units of $\mu$) %$\omega (= \omega/\mu)$  
    for isolated boson stars (left) and dipolar boson stars (right), for five values of $G\in[10^{-3},10^{-2}]$ (in units of $\Phi_0^{-2}$). The horizontal black lines mark three representative values, $Q=40,\,60,\,80$ (corresponding to $Q_\text{dipole}=80,\,120,\,160$ for the dipolar configurations), which we use to construct sequences of solutions with varying $G$.}
    \label{fig:om_Q}
\end{figure*}

\begin{table*}
    \centering
    \renewcommand{\arraystretch}{1.3}
    \begin{tabular}{|c|c||c|c|c|c|c|c|c||c|c|c|}
        \hline
        & & \multicolumn{7}{c||}{Isolated BS} &
        \multicolumn{3}{c|}{Dipolar BS} \\
        \hline
        $~~~Q~~~$ & $G$ & $~~~~\omega~~~~$ & $~~~~M~~~~$ & $~~~~\mathcal{C}~~~~$ & $~~~~q_s~~~~$ &
        $k_2^E$ & $k_3^E$ & $k_2^B$ & $~~~~\omega~~~~$ & $~~~~M~~~~$ & $~~~~L_z~~~~$ \\
        \hline
        
        $40$ & $1\times 10^{-3}$ & $0.8510$ & $40.80$ &
        $0.0068$ & $2.638$ &
        $2.602\times 10^{7}$ & $5.307\times 10^{11}$ &
        $-2.118\times 10^{6}$ & $0.8476$ & $81.48$ & $13.30$ \\

        $60$ & $1\times 10^{-3}$ & $0.7443$ & $56.59$ &
        $0.0105$ & $5.485$ &
        $4.412\times 10^{6}$ & $2.239\times 10^{10}$ &
        $-4.117\times 10^{5}$ & $0.7407$ & $112.9$ & $11.71$ \\

        $80$ & $1\times 10^{-3}$ & $0.6880$ & $70.87$ &
        $0.0132$ & $8.799$ &
        $1.827\times 10^{6}$ & $5.454\times 10^{9}$ &
        $-1.834\times 10^{5}$ & $0.6840$ & $141.3$ & $11.47$ \\

        \hline
        
        $40$ & $1\times 10^{-2}$ & $0.6710$ & $36.73$ &
        $0.0825$ & $3.077$ &
        $646.2$ & $5.337\times 10^{4}$ &
        $-79.69$ & $0.6232$ & $71.47$ & $6.682$ \\
        
        $60$ & $1\times 10^{-2}$ & $0.5676$ & $49.02$ &
        $0.1161$ & $5.710$ &
        $125.1$ & $4.560\times 10^{3}$ &
        $-20.44$ & $0.4990$ & $93.78$ & $5.778$ \\
        
        $80$ & $1\times 10^{-2}$ & $0.5025$ & $59.68$ &
        $0.1440$ & $8.665$ &
        $40.74$ & $884.5$ &
        $-8.527$ & - & - & - \\
        \hline
    \end{tabular}
    \caption{Physical quantities characterizing the isolated and dipolar boson star configurations. The dipolar configurations reported here carry twice the Noether charge of the corresponding isolated boson star, $Q_\text{dipole}=2Q$. Note that for $Q=80$ and $G=10^{-2}$, no dipolar boson star solution exists.}
    \label{tab:quant_bs}
\end{table*}

This control over $Q$ allows us to compare isolated and dipolar boson stars at fixed Noether charge (in other words, at fixed particle number): namely, we compare isolated boson stars of Noether charge $Q$ to  dipolar boson stars of total Noether charge $Q_\text{dipole}=2Q$. The latter are then viewed as bound configurations of two identical boson stars, each carrying a charge $Q$, held in static equilibrium. In this picture, the difference between the mass of the dipolar configuration and twice the mass of the isolated star {measures the gravitational binding energy of the system,
which will be our primary physical variable.} %\rg{Our primary variable is rather $\gamma$ which is different from $M_\text{dipole}-2M$. I would remove the end of the sentence which can be misleading.}

In Fig.~\ref{fig:dbs_vs_ibs}, we compare the scalar profile of an isolated boson star with $Q=60$ to that of a dipolar boson star with $Q_\text{dipole}=120$. The profiles are shown along the $z$-axis, with $z=r\cos\theta$, for two values of the gravitational coupling $G$. In the \textit{weak-gravity} regime, $G=10^{-3}$, the two scalar lumps in the dipolar configuration are sufficiently far apart that the scalar profile in the upper hemisphere ($z>0$) closely matches that of the isolated boson star. The scalar interaction between the two lumps exists, preventing them from collapsing, but is weak: the dipole profile can be well approximated by a simple superposition of two isolated star profiles, $\phi_\text{dipole}\simeq\phi(z-z_0)-\phi(z+z_0)$, where $z_0$ denotes the location of the maximum of $\phi_\text{dipole}$ along the $z$-axis. By contrast, for $G=10^{-2}$, the two lumps are quite close to each other and overlap significantly, so that the exponential tail of the isolated boson star profile no longer provides a good approximation in the region between them. %Consequently, the dipolar profile for $z>0$ differs significantly from that of the isolated boson star. 
We refer to this regime as \textit{strong gravity}. 

\begin{figure*}[t]
    \centering
    %\begin{minipage}{0.48\textwidth}\centering
    %    \includegraphics[width=\textwidth]{G_M.pdf}\%\{\small (a)}
    %\end{minipage}\hfill
    %\begin{minipage}{0.48\textwidth}\centering
    %    \includegraphics[width=\textwidth]{G_om.pdf}\\{\small (b)}
    %\end{minipage}\\[3mm]
    %\begin{minipage}{0.48\textwidth}\centering
    %    \includegraphics[width=\textwidth]{G_r99_dist.pdf}\\{\small (c)}
    %\end{minipage}\hfill
    %\begin{minipage}{0.48\textwidth}\centering
    %    \includegraphics[width=\textwidth]{G_qs.pdf}\\{\small (d)}
    %\end{minipage}
    \includegraphics[width=0.49\textwidth]{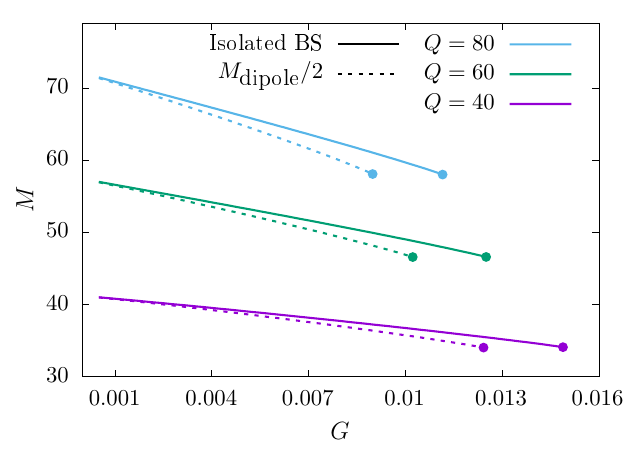}\hfill
    \includegraphics[width=0.49\textwidth]{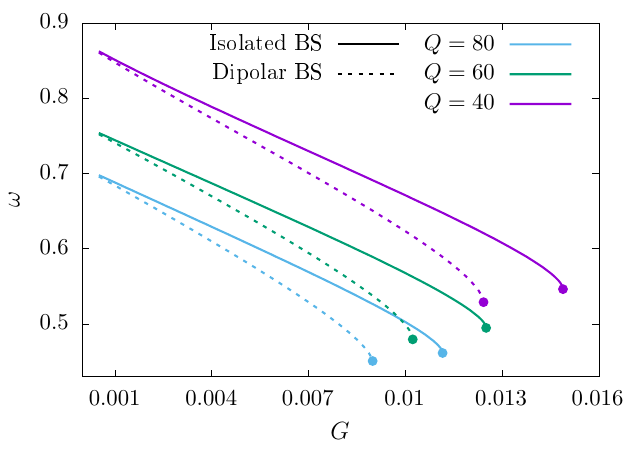}\\
    \includegraphics[width=0.49\textwidth]{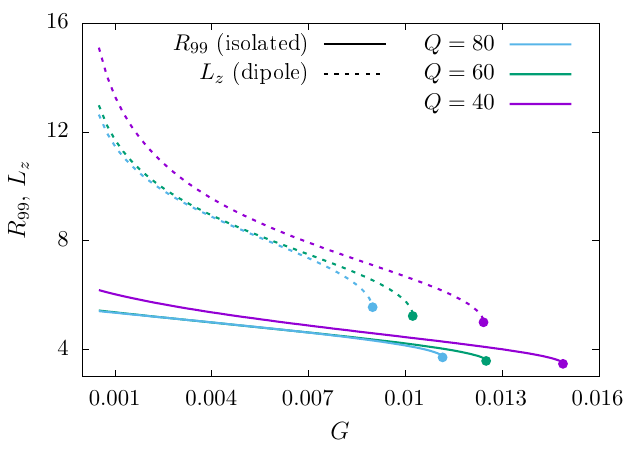}\hfill
    \includegraphics[width=0.49\textwidth]{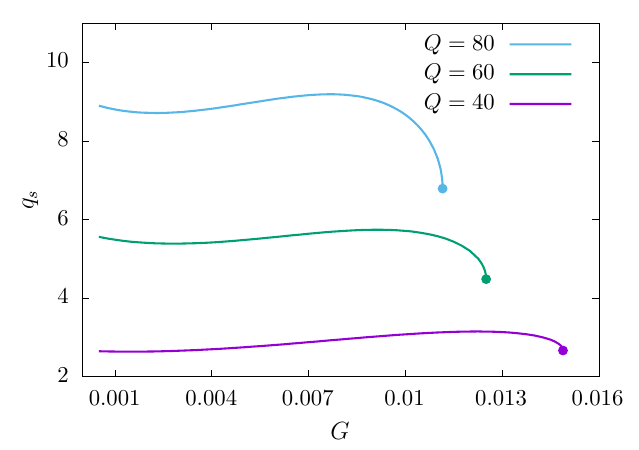}
    \caption{\small Dependence on the gravitational coupling $G$ of various quantities characterizing isolated and dipolar boson stars, for the Noether charge values $Q=40,\,60,\,80$ (corresponding to $Q_\text{dipole}=80,\,120,\,160$). Dots mark the end of each solution sequence (no solutions are found for larger $G$). \textit{Top left:} mass $M$ of the isolated stars, compared to half the mass of the corresponding dipoles. \textit{Top right:} scalar-field frequency $\omega$ for the isolated and dipolar configurations. \textit{Bottom left:} effective radius $R_{99}$ of the isolated boson stars, compared to the proper separation $L_z$ between the two lumps of the dipolar configurations. \textit{Bottom right:} scalar charge $q_s$ of the isolated boson stars.}
    \label{fig:quant_G}
\end{figure*}

Both isolated and dipolar boson stars exhibit a similar domain of existence, usually represented in $(\omega,M)$ diagrams, see \textit{e.g.} Ref.~\cite{Ildefonso2023}. Here, instead, we show the families of solutions in $(\omega,Q)$ diagrams in Fig.~\ref{fig:om_Q}. The solutions emerge from the trivial vacuum at $\omega=1$ (i.e. $\omega=\mu$ in physical units) and exist for $\omega<1$. When $1- \omega \ll 1$ the solutions are very dilute, and as $\omega$ decreases, $Q$ (and $M$) rapidly increases up to a local maximum (close to $\omega=1$ for small $G$). The regime between this maximum and $\omega=1$ is called the \textit{Newtonian branch}. The Noether charge then decreases to a local minimum before increasing again up to the global maximum\footnote{In Fig.~\ref{fig:om_Q}, the global maximum lies outside the vertical-axis range for $G=0.001,\,0.002,\,0.0045$.}. The region between the local minimum and the global maximum is called the \textit{relativistic branch}, as the solutions there are typically more compact. Boson stars are dynamically stable on both the Newtonian and relativistic branches, whereas the intermediate branch is unstable \cite{Tamaki2011,Ildefonso2023}. The families of solutions extend beyond the relativistic branch, down to a minimal frequency, whose value depends on $G$, below which no solutions are found\footnote{In principle, the minimal frequency corresponds to a back-bending point, beyond which the families could be continued toward larger values of $\omega$. However, all solutions beyond the relativistic branch are unstable and are therefore not considered in this work.}. 

Let us now focus on the relativistic branch of isolated boson stars and select three reference charge values, $Q=40,\,60,\,80$, along with the corresponding dipolar configurations with $Q_\text{dipole}=80,\,120,\,160$. For each of these fixed Noether charge values, we vary the gravitational coupling $G$,  %$G (= G\Phi_0^2)$, 
thereby scanning the parameter space along the black horizontal lines shown in Fig.~\ref{fig:om_Q}. The resulting solution sequences will be used for comparison with the analytical model of Sec.~\ref{sec:eob_model}, describing the static two-body interaction. For now, we present various physical quantities of interest obtained numerically.

%Various physical quantities characterizing the isolated and dipolar boson stars are reported in Table~\ref{tab:quant_bs}. 
In addition to the mass $M$ and scalar-field frequency~$\omega$, we report in Table~\ref{tab:quant_bs} {three physical characteristics of isolated boson stars:} (i) the compactness $\mathcal{C}=GM/R_{99}$, where $R_{99}$ is the effective areal radius enclosing 99\% of the total mass; (ii) the \textit{scalar charge} $q_s$ ~\cite{Damour:2025oys} (not to be confused with the Noether charge $Q$), extracted from the asymptotic behavior of the scalar field around an isolated boson star:
\begin{equation}
    \phi(r\to\infty)\approx\frac{q_s}{r}\,\exp\left[-kr+GM\frac{2\omega^2-\mu^2}{k}\log(2e^{\gamma_\textrm{E}}kr)\right],
\end{equation}
where $k=\sqrt{\mu^2-\omega^2}$ and $\gamma_\textrm{E}\approx 0.5772$ is Euler's constant; and (iii) the $\ell=2,3$ electric- and $\ell=2$ magnetic-type tidal Love numbers\footnote{We normalize the tidal Love numbers as in Eq.~(3) of Ref.~\cite{Cardoso:2017cfl}} $k_2^E$, $k_3^E$, and $k_2^B$, obtained by solving the static perturbation equations of an isolated boson star \cite{Damour2009,Cardoso:2017cfl,Sennett:2017etc}. %\td{Our Love numbers are normalized as in Eq. (3) of \cite{Cardoso:2017cfl}. ???}
For dipolar boson stars, we also report the proper separation $L_z$ between the two scalar lumps, defined as
\begin{equation}
    L_z \equiv 2\int_0^{z_0}dr\,e^{F_1(r,0)}.
\end{equation}

The physical quantities for the sequences of solutions with $Q=40,\,60,\,80$ are shown in Fig.~\ref{fig:quant_G}. The gravitational coupling $G$ is varied from $0.5\times 10^{-3}$ up to a maximal value beyond which no solutions are found (marked by the dots in the plots). This maximal value depends on $Q$ and differs between isolated and dipolar boson stars, though for the values of $Q$ considered here it is always of order $10^{-2}$. We start from $G=0.5\times 10^{-3}$ because the numerical construction of dipolar boson stars becomes increasingly challenging as $G\to 0$. We recall that they do not admit a flat-spacetime limit \cite{Cunha:2022tvk}.

In the top left panel, we compare the mass of isolated boson stars with half the mass of the dipoles. One always has $M_\text{dipole}/2<M_\text{isolated}$, reflecting the negative binding energy of the dipolar configurations. In the top right panel, we show the scalar-field frequency $\omega$. Its values for the isolated and dipolar boson stars approach each other in the small-$G$ limit. In the bottom left panel, we compare the radius $R_{99}$ of isolated boson stars with the proper separation $L_z$ of the dipolar configurations. Both quantities increase as $G$ decreases, but $R_{99}$ approaches a finite value for small $G$, where isolated boson stars approach flat-spacetime $Q$-balls~\cite{Coleman:1985ki}, whereas $L_z$ rapidly grows. This behavior of $L_z$ is expected: since gravitational attraction is required to balance the repulsion of the two scalar lumps, the dipole can only be maintained as $G\to 0$ if the lumps move arbitrarily far apart. In this weak-gravity regime, the two lumps therefore barely interact and behave as two essentially isolated boson stars, consistent with $M_\text{dipole}/2$ and $\omega_\text{dipole}$ approaching the isolated-boson-star values as $G\to 0$. Finally, the bottom right panel shows the scalar charge $q_s$ of the isolated boson stars. It exhibits a non-monotonic behavior with respect to $G$, but grows with increasing $Q$.

\section{Analytical Method: An Effective-One-Body Approach}
\label{sec:eob_model}
In this section, we discuss an analytical method describing the dipolar boson stars of Sec.~\ref{sec:numerical}, viewed as two identical boson stars held in static equilibrium by the balance between long-range gravitational attraction and short-range scalar repulsion. The latter arises from the relative phase $\pi$ imposed by the antisymmetry of $\phi$ with respect to the equatorial plane. An simple effective %, phenomenological 
model of this balance was proposed in Ref.~\cite{Cunha:2022tvk}. { Here, we improve the heuristic model of Ref.~\cite{Cunha:2022tvk} in two important ways: (i) following Ref. \cite{Damour:2025oys} we use the concept of scalar charge $q_s$ of each boson star %\rg{To me this sentence is unclear. What do you mean by "\textit{rooted} in each boson star"?} 
(and we will see that the $\pi$ relative phase implies a sign difference $q_2 = - q_1$); and, (ii) we use a recent, accurate version~\cite{Damour:2025uka} of the EOB formalism to describe both the gravitational and the scalar interactions between the two constituents of a dipolar system.}

\subsection{EOB description}
The EOB formalism maps the two-body dynamics onto the motion of an effective particle moving in an effective metric $g_{\mu \nu}^{\rm eff}$~\cite{Buonanno:1998gg, Buonanno:2000ef}. However, this mapping is not unique. Different gauge choices, related by canonical transformations, change the effective metric (or the effective Hamiltonian) while leaving the physical observables invariant. {Depending on the considered physical situation,} a given observable may be better captured in one gauge than in another. 

{%\rg{Check that you are happy with the new notation $M\to M_\text{bin}$ is this paragraph. Using the generic symbol $M$ was a bit ambiguous, I prefer a more explicit notation.} 
A general feature of the EOB formalism is to use as basic energy variable an \textit{effective energy}, $E_{\rm eff}$, given in terms of the real, total energy $E_\text{tot}$ 
of the considered binary system by}
\begin{equation}
    E_{\rm eff} \equiv \frac{E_\text{tot}^2-M_1^2-M_2^2}{2 (M_1+M_2)}\,,
\end{equation}
where $M_1$, $M_2$ denote the individual (rest) masses of the components of the binary. We will work here with the dimensionless effective energy 
\begin{equation} \label{defgamma}
   \gamma \equiv \frac{E_{\rm eff}}{\mu_{\rm eff}} = \frac{E_\text{tot}^2-M_1^2-M_2^2}{2 M_1 M_2}\,,
\end{equation}
where  $\mu_{\rm eff} = M_1 M_2/(M_1+M_2)$ is the usual effective mass of the binary. The (dimensionless) effective energy $\gamma$ contains the same information as the (negative) fractional binding energy $(E_\text{tot}-M_1-M_2)/(M_1+M_2)$. Note that $\gamma<1$ 
for bound binary systems, while $\gamma>1$ for scattering states.

{For bound orbits, the standard post-Newtonian-based EOB uses an energy-independent, Schwarzschild-like gauge~\cite{Buonanno:1998gg, Buonanno:2000ef} which encodes the dynamics in three functions: two metric potentials $A(R) \equiv -g^{\rm eff}_{tt}$, and $B(R) \equiv g^{\rm eff}_{RR}$
(where $R$ is a Schwarzschild radial coordinate), and a Finsler potential $Q_{\rm eff}(R,P_R^2)$ (where $P_R$ is the radial momentum).  For scattering (unbound) orbits, treated within post-Minkowskian gravity, it is often convenient to use an energy-dependent isotropic gauge \cite{Damour:2022ybd} which encodes the dynamics in a \textit{single} energy-dependent effective radial potential $w(\bar{r}, \gamma)$, where $\bar{r}$ denotes the isotropic radial coordinate.  Recently, 
 Ref.~\cite{Damour:2025uka} introduced an energy-dependent, Schwarzschild-like gauge, in which the dynamics is described by a \textit{single} energy-dependent metric potential $A(R,\gamma) \equiv -g^{\rm eff}_{tt}$. The latter Lagrange-EOB (LEOB) description, allows one to combine the knowledge recently acquired from post-Minkowskian gravity, with
 the Schwarzschild-like structure of the post-Newtonian-based EOB which is well adapted to the description of bound systems. This is the approach we shall use here.}

{Reference~\cite{Damour:2025oys} gave the analytical description of the interaction between two boson stars in the $w$-EOB gauge including next-to-leading order (NLO) scalar and tidal contributions, for scattering configurations. In the next sub-section, we show how to transform the NLO scalar-potential results of Ref.~\cite{Damour:2025oys} from the $w$-EOB gauge to the energy-dependent metric potential $A(R,\gamma)$ of a corresponding LEOB gauge.}

\subsection{Relation between $A$-LEOB and $w$-EOB potentials}

In general, two EOB descriptions can be related to each other either by identifying a common gauge-invariant observable (such as the scattering angle $\chi(E,J)$), or  by using a canonical transformation. However, since we are essentially dealing in both cases with a simple (energy-dependent) geodesic action $S=-\mu_{\mathrm{eff}}\int  ds_{\rm LEOB}$,  the $w$-EOB and LEOB descriptions can be connected by a simple, energy-dependent radial-coordinate transformation, $R = f_\gamma(\bar r)$, without the need of a canonical transformation involving both coordinates and momenta. The needed radial-coordinate transformation can be easily constructed by comparing the
two mass-shell conditions. In the following we use dimensionless EOB variables, notably 
$R= R^{\rm phys}/(G (M_1+M_2))$, $\bar r= {\bar r}^{\rm phys}/(G (M_1+M_2))$, 
$P_R= P_R^{\rm phys}/\mu_{\rm eff}$, $p_{\bar{r}}= p_{\bar{r}}^{\rm phys}/\mu_{\rm eff}$, as well as a
dimensionless angular momentum of the system: $j \equiv J^{\rm phys}/(G M_1 M_2)$ (which is the same in
all EOB gauges).

In the $w$-EOB gauge, with the isotropic radial momentum $p_{\bar{r}}$, the mass-shell condition reads
\begin{align}
\label{eq:massshellwEOB}
    p_{\bar{r}}^2 + \frac{j^2}{\bar{r}^2} = \gamma^2 - 1 + w(\bar{r},\gamma)~.
\end{align}

The LEOB gauge uses as effective metric (in the equatorial plane)
\begin{align}
    ds^2_{\rm LEOB} = -A(R, \gamma) dt^2 + B(R, \gamma) dR^2 + R^2 d\varphi^2~,
    \label{eq:metricLEOB}
\end{align}
with $A(R, \gamma) B(R, \gamma) = 1$, so that the corresponding mass-shell condition
$g_{\rm eff}^{\mu \nu} P_{\mu}^{\rm phys} P_{\nu}^{\rm phys} + \mu_{\rm eff}^2=0$ yields
\begin{align}
\label{eq:massshellLEOB}
    P_R^2 = B(R, \gamma) \left[\frac{\gamma^2}{A(R, \gamma)} - 1 -\frac{j^2}{R^2}\right]~.
\end{align}
%A static configuration in equilibrium corresponds to the effective particle at rest, i.e. $P_R = j = 0$, at an equilibrium radius $R_{\ast}$ satisfying 
%\begin{align}
 %   \gamma^2 = A(R_{\ast}, \gamma)\quad \mathrm{and} \quad \partial_R A(R_{\ast}, \gamma) =0~.
%\end{align} 

Comparing \eqref{eq:massshellwEOB} to \eqref{eq:massshellLEOB}, we see that we can transform
the LEOB mass-shell condition into a $w$-EOB equivalent one by transforming, via an energy-dependent radial
change of variables, $R = R(\bar r, \gamma)$, the Schwarzschild-like
LEOB metric \eqref{eq:metricLEOB} into a corresponding isotropic form, namely
\begin{align}
\label{eq:LEOBiso}
    ds^2_{\rm LEOB} = -\bar A(\bar r)\, dt^2 + \bar B(\bar r) \left(d\bar r^2 + \bar r^2 d\varphi^2\right)~.
\end{align}
Indeed, the mass-shell condition corresponding to the isotropic LEOB metric \eqref{eq:LEOBiso} reads
 \begin{align}
\label{eq:massshellLEOBiso}
    p_{\bar{r}}^2 + \frac{j^2}{\bar{r}^2} = \bar B(\bar r)\left[\frac{\gamma^2}{\bar A(\bar r)} - 1\right]~.
\end{align}
Requiring the transformation $R = R(\bar r, \gamma)$ to transform \eqref{eq:metricLEOB} into \eqref{eq:LEOBiso} yields the relations
\begin{subequations}
\label{eq:ABbar}
\begin{align}
    \bar A(\bar r) &= A(R(\bar r))~, \\  
    \bar B(\bar r) &= B(R(\bar r)) \left(\frac{dR}{d\bar r}\right)^2~,  \\ 
    \bar B(\bar r)\, \bar r^2 &= R^2(\bar r)~.
\end{align}
\end{subequations}

Using the (Schwarzschild-gauge) condition $A(R) B(R) = 1$ and eliminating $\bar B$ gives a simple ordinary differential equation (ODE) for the radial map,
\begin{align}
    \frac{d\bar r}{\bar r} = \frac{dR}{R}\sqrt{B(R)} = \frac{dR}{R}\frac{1}{\sqrt{A(R)}}~.
    \label{eq:drr}
\end{align}
In addition, given the knowledge of the radial map $R(\bar{r}, \gamma)$, equating the right-hand side of the isotropic-gauge LEOB mass-shell condition \eqref{eq:massshellLEOBiso} with that of the $w$-EOB mass-shell condition \eqref{eq:massshellwEOB} gives the following exact link between the $A (R,\gamma)$  and $w (\bar{r},\gamma)$ potentials:
\begin{align}
    \bar r^2 \left[\gamma^2 - 1 + w(\bar r,\gamma)\right] = R(\bar{r})^2 \left[\frac{\gamma^2}{A(R(\bar{r}))} - 1\right]~.
    \label{eq:exact}
\end{align}
Considered jointly, the two equations~\eqref{eq:drr} and~\eqref{eq:exact} give an algebraico-differential system allowing one, if $w(\bar r,\gamma)$ is known, to determine both the radial map and $A(R,\gamma)$. 

To obtain the corrections to the $A(R)$ potential due to the scalar interactions, we solve the above equations perturbatively by setting
\begin{align}
    A(R,\gamma)& = A_0(R,\gamma) + \epsilon A_1(R,\gamma)+ O(\epsilon^2)~,\\
    w(\bar r,\gamma) & = w_0(\bar r,\gamma) + \epsilon w_1(\bar r,\gamma)+ O(\epsilon^2)~,
\end{align} 
where $w_0$ and $A_0$ describe the known point-particle (or black hole) contributions and $\epsilon$ is a book-keeping parameter for the scalar contributions (including the boson-star gravitational tidal contributions (see Sec. \ref{sec:num_vs_analytic})). Correspondingly, the radial map can be written as
\begin{align}
    \bar r = \bar r_0(R,\gamma) + \epsilon \bar r_1(R,\gamma) + O(\epsilon^2)~.
    \label{eq:rexp}
\end{align}
Our problem is to transform our current (NLO) knowledge of $w_1(\bar r,\gamma)$~\cite{Damour:2025oys} into the knowledge of the corresponding $A_1(R,\gamma)$.

Inserting the expansions above into Eq.~\eqref{eq:drr}, we obtain at zeroth order $\ln \bar r_0(R) = \int dR/(R\sqrt{A_0}) + {\rm const}$, and at first order
\begin{align}
    \frac{d}{dR}\left(\frac{\bar r_1}{\bar r_0}\right) = -\frac{1}{2}\frac{A_1(R)}{R A_0^{3/2}(R)}~.
    \label{eq:R2}
\end{align}
Similarly, the first-order terms of Eq.~\eqref{eq:exact} give
\begin{align}
    2\bar r_0 \bar r_1 \left[\gamma^2 - 1 + w_0(\bar r_0)\right] &+ \bar r_0^2 \left[w_1(\bar r_0) + \bar r_1 \frac{dw_0}{d\bar r_0}\right] \nonumber\\
    &= -\frac{\gamma^2 R^2}{A_0^2} A_1(R)~.
    \label{eq:R3}
\end{align}
Defining 
\begin{equation}
    \rho(R) \equiv \frac{\bar r_1(R)}{\bar r_0(R)}
\end{equation}
 and the zeroth-order quantity
\begin{align}
    \mathcal{J}_0 \equiv \bar r_0^2\left(\gamma^2 - 1 + w_0\right) = R^2\left(\frac{\gamma^2}{A_0} - 1\right)~,
\end{align}
Eq.~\eqref{eq:R3} can be written as
\begin{align}
    A_1(R) = -\frac{A_0^2}{\gamma^2 R^2}\left[\bar r_1 \frac{d\mathcal{J}_0}{d\bar r_0} + \bar r_0^2\, w_1(\bar r_0)\right]~.
\end{align}
Using $d\mathcal{J}_0/d\bar r_0 = (R\sqrt{A_0}/\bar r_0)\, d\mathcal{J}_0/dR$, which follows from the zeroth order of Eq.~\eqref{eq:drr}, we obtain
\begin{align}
    A_1(R) = -\frac{A_0^2 \bar r_0^2}{\gamma^2 R^2}\left[\rho\, \frac{R\sqrt{A_0}}{\bar r_0^2}\frac{d\mathcal{J}_0}{dR} + w_1(\bar r_0)\right]~.
    \label{eq:A1rho}
\end{align}
Inserting Eq.~\eqref{eq:A1rho} into Eq.~\eqref{eq:R2} gives a first-order linear ODE for 
the function $\rho(R)$,
\begin{align}
    \frac{d\rho}{dR} = \frac{A_0}{2\gamma^2 R^2}\frac{d\mathcal{J}_0}{dR}\, \rho + \frac{\sqrt{A_0}\, \bar r_0^2}{2\gamma^2 R^3}\, w_1(\bar r_0)~.
    \label{eq:rhoODE}
\end{align}
Given the function $w_1(\bar r_0)$, the algebraic-differential system of equations \eqref{eq:A1rho}, \eqref{eq:rhoODE} determines the looked-for function $A_1(R)$.

The coefficients with the subscript ``${}_0$" entering the latter equations are fixed by the known point-particle dynamics, while the function $w_1(\bar r_0)$ is known (at NLO level) from Ref.~\cite{Damour:2025oys} (see \eqref{w1} below). In view of the fractional first-post-Minkowskian knowledge of $w_1(\bar r_0)$, it is enough to integrate the system \eqref{eq:A1rho}, \eqref{eq:rhoODE}, 
by using only a corresponding first-post-Minkowskian accuracy in $A_0(R)$, i.e. $A_0 = 1 - 2/R$,
leading to $\bar r_0(R)$ defined by solving $R = \bar r_0\,(1 + 1/(2\bar r_0))^2$.

We solve (for each given value of  $\gamma$) Eq.~\eqref{eq:rhoODE} with the boundary condition $\rho \to 0$ as $R \to \infty$, since the short-range scalar interactions become negligible at large separations. We then insert the result into Eq.~\eqref{eq:A1rho} to obtain the NLO LEOB potential $A_1(R,\gamma)$ corresponding to $w_1(\bar r_0)$. 
{In view of the analytical complexity of the system of equations \eqref{eq:A1rho}, \eqref{eq:rhoODE} we determine the function $A_1(R,\gamma)$ by numerical integration.}

\section{Numerical vs analytical comparison}
\label{sec:num_vs_analytic}

We now compare our analytical prediction with three numerical sequences of dipolar boson star configurations with individual Noether charges  %$Q_1=Q_2=40,\,60,\,80$, 
$Q=40,\,60,\,80$ and varying  gravitational coupling $G$. 
%We now compare our analytical prediction with the numerical sequences of dipolar boson star configurations, with the Noether charge $Q = 40,~60,~80$ for each isolated star, obtained by varying the gravitational constant $G$. 
%The analytically predicted effective energy $\gamma$ is defined by $\gamma^2 = A(R_\ast,\gamma)$ and $\partial_R A(R_{\ast}, \gamma) =0$ with the 
Our analytical model is an LEOB $A(R,\gamma)$  potential given by the sum
\begin{align}
    A(R,\gamma) = A^{\rm BH}(R,\gamma) + A^{\rm scalar}(R,\gamma)~,
    \label{eq:Asplit}
\end{align}
where the point-mass (or black hole) contribution is taken at the third post-Minkowskian accuracy, i.e. (denoting $h=\sqrt{1+ 2 \nu (\gamma-1)}$)
\begin{equation} \label{ABH}
  A^{\rm BH}(R, \gamma)=   1 - \frac{2}{R} + 
		 \frac{6 \nu}{R^2 } \frac{(\gamma-1) (5 \gamma ^2-1)}{ h(h+1) (3 \gamma ^2-1)}+ \frac{a_3(\gamma)}{R^3}\,,
\end{equation}
with the $a_3(\gamma)$ coefficient given in Eq. (2.69a) of \cite{Damour:2025uka}, and where
$A^{\rm scalar} \equiv A_1$ is the scalar contribution obtained from $w_1$ through Eqs.~\eqref{eq:rhoODE} and~\eqref{eq:A1rho}. The $w_1(\bar r)$ scalar potential for two boson stars reads~\cite{Damour:2025oys} 
%\rg{For the moment, I put the equation below in a widetext environment, to avoid using $e^{\dots}$ with a large argument. Using $\exp(\dots)$ is easier to read.}
%\begin{align}
%    w_1= & -\frac{8 \pi c_1 c_2}{G m_1 m_2} \frac{e^{-\bar{m} \bar{r}}}{\bar{r}} e^{G M ((2\omega^2-\mu^2)/\widetilde{\mu})  \log(2G M e^{\gamma_{\rm E}} \widetilde{\mu}{\bar{r}})}\,.
%\end{align}
\begin{widetext}
\begin{equation} \label{w1}
    w_1(\bar r)=+\frac{8\pi q_1 q_2}{GM_1 M_2}\frac{e^{-\bar{m} \bar{r}}}{\bar{r}}\exp\left(\frac{GM_{\rm tot}(2\omega^2-\mu^2)}{\widetilde{\mu}}\log(2G M_{\rm tot} e^{\gamma_{\rm E}} \widetilde{\mu}{\bar{r}})\right).
\end{equation}
\end{widetext}
Here, the $\pi$ relative phase of dipolar configurations induces opposite scalar charges $q_1=- q_2$ (yielding repulsion rather than attraction) of the isolated boson stars of mass $M_1=M_2$ (and total rest mass $M_{\rm tot}=M_1+M_2$), $\omega(=\omega_1 =\omega_2)$ is their scalar-field frequency, and $\bar{m} \equiv GM_{\rm tot} h \widetilde{\mu}$ [with $\widetilde{\mu}^2 \equiv \mu^2- (2
\omega_1\omega_2\gamma-\omega_1^2-\omega_2^2)/(\gamma^2-1)$] is the dimensionless inverse range of the interaction. %, and $\gamma_E$ is Euler's constant.  
For each value of $G$ along a given sequence, the mass $M_1=M_2$, the frequency $\omega=\omega_1 =\omega_2$, and the scalar charge $q_1=- q_2$ are taken from our numerical solutions for isolated boson stars with the same Noether charges as the individual lumps in the dipole. %However, we do not take the effective energy $\gamma$ from the numerical solutions, but determine it self-consistently by bootstrapping the equilibrium condition

{We recall that the mass-shell condition in LEOB gauge says that the effective energy $\gamma$ of a binary system is related to the orbital angular momentum $j$, and the (relative) radial momentum, $P_R$, by}
\begin{equation}
    \gamma^2= A(R, \gamma) \left(1+ \frac{j^2}{R^2} + A(R, \gamma) P_R^2 \right)\,.
\end{equation}
{The right-hand side of the latter equation is naturally interpreted as the sum of a potential-energy term, $A(R, \gamma)$, a centrifugal term $\propto j^2/R^2$, and a radial kinetic-energy term $\propto P_R^2$.}

{In the present work, we are considering binary systems with zero angular momentum, and we look for equilibrium solutions with vanishing radial kinetic energy. Such solutions clearly correspond to minimizing the potential energy 
contribution $A(R, \gamma)$ with respect to the radial variable $R$. However, as our radial potential $A(R, \gamma)$ also depends on $\gamma$, the 
analytical effective energy must be self-consistently determined by 
 bootstrapping the latter equilibrium condition, i.e. by finding solutions of}
\begin{align}
    \gamma^2 = \min_R A(R,\gamma)~.
    \label{eq:gammaAR}
\end{align}
{This is equivalent to finding solutions of the two equations $\gamma^2 = A(R_\ast,\gamma)$ and $\partial_R A(R_\ast,\gamma) = 0$, together with the inequality $\partial_{RR} A(R_\ast,\gamma) > 0$.} We solve Eq.~\eqref{eq:gammaAR} for $\gamma$ with a root-finding algorithm. 
Given values of $Q$ and $G$, our analytical equilibrium condition \eqref{eq:gammaAR} determines
at most one solution, say $\gamma_\text{analytic}$.

\begin{figure*}[t]
    \centering
    \begin{minipage}{0.48\textwidth}\centering
        \includegraphics[width=\textwidth]{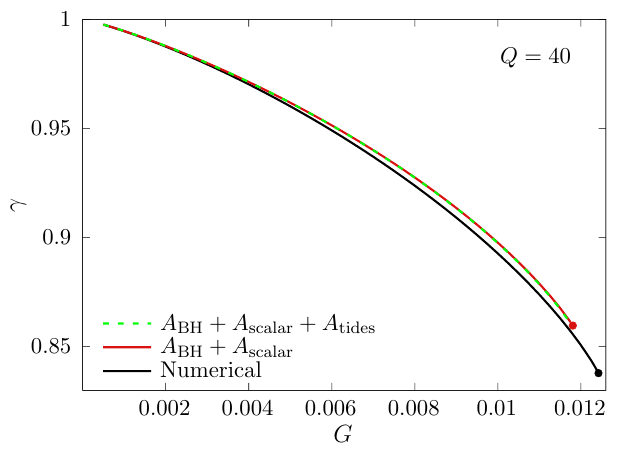}\\{\small (a)}
    \end{minipage}\hfill
    \begin{minipage}{0.48\textwidth}\centering
        \includegraphics[width=\textwidth]{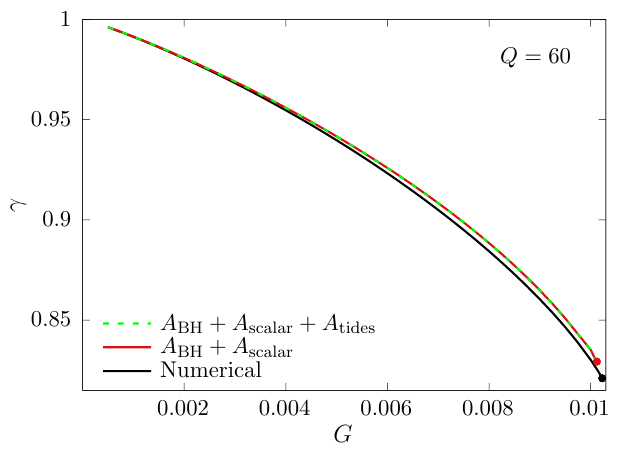}\\{\small (b)}
    \end{minipage}\\[3mm]
    \begin{minipage}{0.48\textwidth}\centering
        \includegraphics[width=\textwidth]{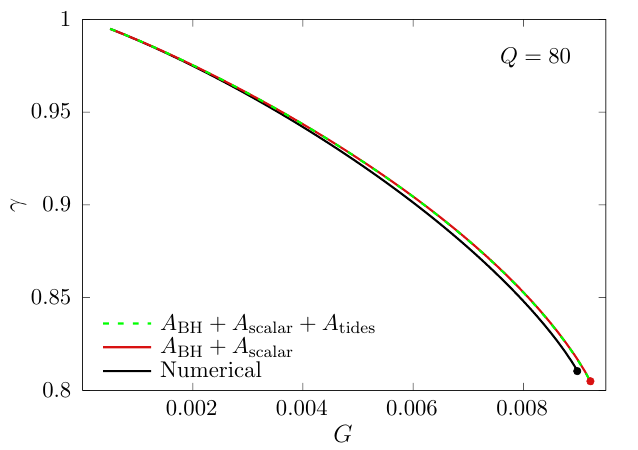}\\{\small (c)}
    \end{minipage}\hfill
    \begin{minipage}{0.48\textwidth}\centering
        \includegraphics[width=\textwidth]{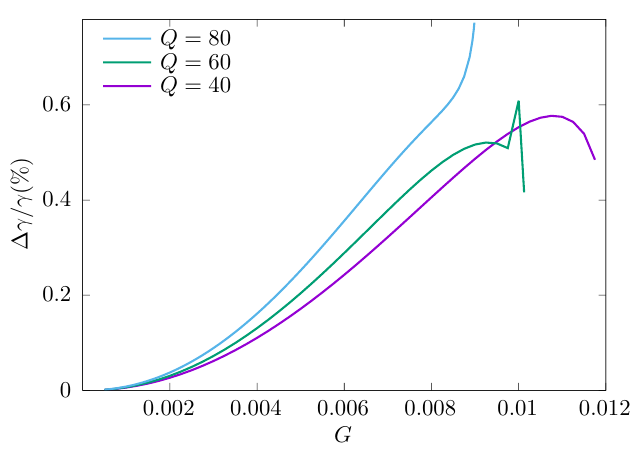}\\{\small (d)}
    \end{minipage}
    \caption{\small (a)--(c) Comparison of the effective energy $\gamma$ of static dipolar boson stars as a function of the gravitational coupling $G$, for Noether charge $Q = 40,~60,~80$ of the individual boson star constituents. The numerical results are shown as a black solid line, and the analytical results, obtained by bootstrapping the effective energy, as a red solid line. The green dashed line shows the analytical results including tidal contributions. The black dot marks the last numerical dipolar boson star configuration, and the red dot represents the turning point beyond which the analytical model predicts no static equilibrium. (d) Fractional difference ($\Delta \gamma /\gamma$) in \% between the numerical and analytical values (without tides) of $\gamma$ as a function of $G$, for $Q = 40,~60,~80$.}
    \label{fig:NRvsAR}
\end{figure*}

The analytical prediction $\gamma_\text{analytic}(G,Q)$ (when it exists) is then compared
with the corresponding numerical value $\gamma_\text{numerical}(G,Q)$ %which is given by its definition \eqref{defgamma}, i.e.
which, using Eq.~\eqref{defgamma}, is given by,
\begin{equation}
    \gamma_\text{numerical}(G,Q)=\frac{1}{2}\left(\frac{M_\text{dipole}}{M_\text{isolated}}\right)^2-1,
\end{equation}
where $M_\text{dipole}$ is the mass of the dipolar configuration, and $M_\text{isolated}$ is the mass of the isolated boson star carrying the same Noether charge $Q$ as each of the two lumps composing the dipole. The results for Noether charges $Q = 40,~60,~80$ are shown in Figs.~\ref{fig:NRvsAR} (a--c), where the numerical values of $\gamma$ (black solid line) are compared with the analytical predictions (red solid lines). For all three sequences, the analytical predictions agree very well (given our limited analytical accuracy) with the numerical results over the whole range of $G$, and for the three values of $Q$ we have selected. In Fig.~\ref{fig:NRvsAR} (d), we show the fractional difference 
$\Delta\gamma/\gamma = \gamma_{\rm analytic}/\gamma_{\rm numerical}-1$  in percent.  Overall, the deviation grows with $G$ and remains below $0.77$\% for all three sequences.

As $G$ increases along a sequence, the gravitational attraction between the two stars grows relative to their scalar repulsion. {As already highlighted in Sec.~\ref{sec:numerical},} beyond a critical value of $G$, the scalar repulsion can no longer balance the gravitational pull, and no stable equilibrium exists, i.e., we do not find dipolar boson star solutions in this regime. In the analytical model, this corresponds to the disappearance of the minimum of the LEOB potential. As $G$ increases, one indeed finds that the minimum of $A$ becomes shallower and eventually merges with a neighboring maximum, thereby creating an inflection point and no minimum. %(see Fig.~\ref{fig:Vpot} for the $Q=60$ sequence). 
At this critical configuration, the function $F(R) \equiv \gamma^2 - A(R,\gamma)$ has a triple zero:
\begin{align}
    F = 0~, \qquad \partial_R F = 0~, \qquad \partial_R^2 F = 0~.
    \label{eq:critical}
\end{align}
For a given $Q$, the latter three equations determine a critical radius $R_{\rm c}(Q)$, a critical effective energy $\gamma_{\rm c}(Q)$, and a critical gravitational coupling $G_{\rm c}(Q)$. For $G > G_{\rm c}(Q)$, the LEOB potential admits no static equilibrium, and the analytical description predicts no dipolar boson star configuration. These critical configurations are shown as red solid dots in Fig.~\ref{fig:NRvsAR} (a--c). For all three sequences, we find that the analytical critical coupling $G_{\rm c}(Q)$ lies {close to, but} slightly away from the value at which the numerical sequence ends. %\rg{I would mention that the critical value of $G$ predicted by the analytical model is different from the numerical results: according to the figure, the analytical $G_c$ is always slightly below the numerical $G_c$. Apparently, the discrepancy between the two is smaller when $Q$ increases. }
\begin{figure}[t]
    \centering
    \begin{minipage}{0.48\textwidth}\centering
        \includegraphics[width=\textwidth]{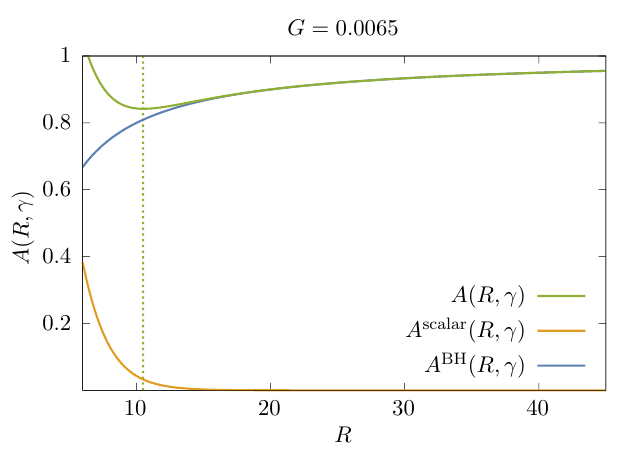}\\{\small (a)}
    \end{minipage}\hfill
    \begin{minipage}{0.48\textwidth}\centering
        \includegraphics[width=\textwidth]{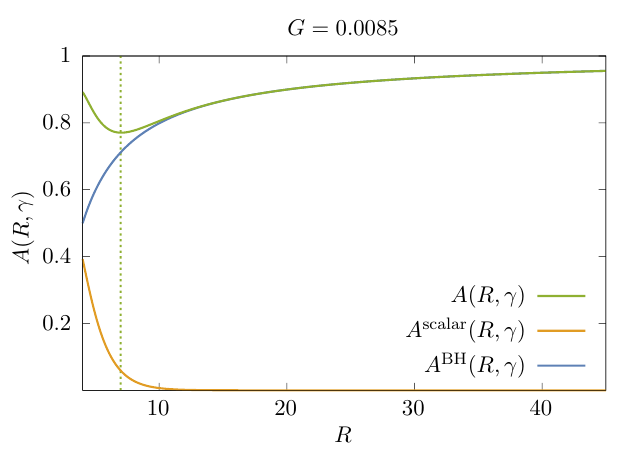}\\{\small (b)}
    \end{minipage}\hfill
    \caption{\small
 LEOB potential $A(R,\gamma)$ (green solid line) for the $Q = 60$ sequence, evaluated at the bootstrapped analytical value (without tides) of the effective energy $\gamma$, for (a) $G = 0.0065$ ($\gamma^{\rm analytic} = 0.917$) and (b) $G = 0.0085$ ($\gamma^{\rm analytic} = 0.877$). Also shown are its two building blocks, the attractive 3PM Schwarzschild-like gravitational potential $A^{\rm BH}(R,\gamma)$ (blue solid line) and the repulsive Yukawa-like scalar potential $A^{\rm scalar}(R,\gamma)$ (orange solid line). Their combination produces a local minimum of $A$ at $R = R_\ast$ (green dashed line), where $A(R_\ast,\gamma) = \gamma^2$, corresponding to the static equilibrium. 
}
    \label{fig:APotential}
\end{figure}

We also include the tidal contributions $w_6^{\rm tid}/\bar r^6 + w_7^{\rm tid}/\bar r^7$ (see Eq.~(8) of Ref.~\cite{Damour:2025oys}) due to {both ($\ell=2$) electric-type and ($\ell=2$) magnetic-type Love numbers} 
in $w_1$, in the same way as the scalar contribution, and find that their effect on $\gamma_{\rm analytic}$ is negligible. The tidal results are shown as green dashed lines in  Fig.~\ref{fig:NRvsAR} (a--c). We find that the tidal effects are consistently less important for these static configurations than in the scattering of boson stars, where they were found to improve the agreement with numerical simulations at small impact parameters~\cite{Damour:2025oys}.%\tj{More negligible than scattering}\rg{ We could also compare the role of tides in this equilibrium study to that in your scattering paper. It seems like the tidal effects are even more negligible in these static configurations than in the dynamical scattering.}

%The binding energy (equivalent to $\gamma$) 
The effective energy $\gamma$ (related to the binding energy of the dipole) is the only gauge invariant variable that makes sense both in our numerical approach and in the EOB approach, for the systems considered here.
By contrast, the equilibrium EOB radial variable $R_\ast(G,Q)$ is not gauge invariant and therefore cannot be meaningfully \textit{quantitatively} compared with the numerical equilibrium proper distance $L_z(G,Q)$. Nevertheless, it is instructive to compare the two to see if they are
in \textit{qualitative} agreement. We found that they {are in reasonable qualitative agreement}. Let us give a few examples of the comparison between $R_\ast(G,Q)$ and $L_z(G,Q)$ along with the structure of the potential $A(R,\gamma)$. 

%\rg{The transition between these two paragraphs is a bit abrupt. We say we will compare $R_\ast$ with $L_z$ but then we discuss Fig.5 and the structure of the potential $A(R,\gamma)$.}

Figure~\ref{fig:APotential} compares the potential $A(R,\gamma)$ in the two cases $G=0.0065,~0.0085$ (for $Q=60$) by displaying not only the complete %corresponding LEOB 
potential but also its two building blocks $A^{\rm BH}{(R, \gamma)}$ and $A^{\rm scalar}(R, \gamma)$, see Eq. \eqref{eq:Asplit}.
The figure explicitly shows how the attractive third-post-Minkowskian Schwarzschild-like gravitational potential $A^{\rm BH}{(R, \gamma)}$ (blue solid line) combines with the repulsive Yukawa-like scalar potential 
(orange solid line) to create a local minimum in the total potential $A(R, \gamma)$  (green solid line) at $R_{\ast}$ (green dashed line). In the examples shown, the radius (self-consistently) minimizing the total $A(R, \gamma)$
is $R_\ast \approx 10.508$ for $G = 0.0065$ and $R_\ast \approx 6.977$ for $G = 0.0085$, corresponding to the bootstrapped values $\gamma^{\rm analytic}= 0.917377$ and $\gamma^{\rm analytic}= 0.87723$, respectively. These are quite close to their corresponding numerical values $\gamma^{\rm numerical}= 0.914328$ and $\gamma^{\rm numerical}= 0.872908$. In order to compare the dimensionless Schwarzschild-like radial coordinates $R_\ast \approx 10.508$ and $R_\ast \approx 6.977$ to the corresponding numerical proper distances $L_z = 6.086$ and $L_z = 4.753$, we need to recall that $R \equiv R^{\rm phys}/( G (M_1+M_2))$ while $L_z = \mu L_z^{\rm phys}$, so that 
$R^{\rm phys}/L_z^{\rm phys}= G (M_1+M_2) \mu \times R/L_z$. In the present case, the conversion factor $K \equiv G (M_1+M_2) \mu$ is $K \approx 0.67689$ and $K \approx 0.85596$, respectively. {Therefore the equilibrium radii displayed in Fig. \ref{fig:APotential} corresponds (in units of $\mu^{-1}$) to
$K R_\ast^{\rm analytic} \approx 7.113$ and $K R_\ast^{\rm analytic} \approx 5.972$, respectively, which are reasonably close to 
$L_z = 6.086$ and $L_z = 4.754$. In Table \ref{tab:results} we give a few more examples of the comparison between
$K R_\ast^{\rm analytic}$  and $L_z^{\rm numerical}$. In view of the lack of gauge-invariance of $R_{\ast}$, this comparison can only be qualitative, and we consider the level of agreement shown in Table~\ref{tab:results} to be reasonable.} %\rg{I would remove the end of the sentence and just say "In view of the lack of gauge-invariance of $R_{\ast}$, this comparison can only be qualitative." (Saying "reasonably satisfactory" is very vague.)}

\begin{table}[t]
\centering
\caption{Comparison between the bootstrapped analytical and numerical values of the effective energy $\gamma$, and between the EOB equilibrium radius $R_\ast$ and the numerical proper separation $L_z$ of the two lumps, for $Q = 40,~60,~80$. Here $K \equiv G M_{\rm tot}\,\mu$ converts $R_\ast$ (in units of $G M_{\rm tot}$) into  $L_z$ (in units of $1/\mu$).}
\label{tab:results}
\begin{ruledtabular}
\begin{tabular}{ccccccc}
$Q$ & $G$ & $K$ & $\gamma^{\rm numerical}$ & $\gamma^{\rm analytic}$ & $L_z$ & $K R_\ast$  \\
\hline
\multirow{2}{*}{40} & 0.001  & 0.08160  & 0.994583  & 0.994641  & 13.0534  & 13.3217 \\
                    & 0.003  & 0.23975  & 0.979366  & 0.979969  & 9.75697  & 10.2903\\
                    & 0.0065 & 0.49922  & 0.943195  & 0.945853  & 6.89321  & 7.80928 \\
                    & 0.009  & 0.67002  & 0.909096  & 0.913522  & 5.43847  & 6.45993 \\
                    & 0.011  & 0.79715  & 0.873806  & 0.878831  & 4.2972   & 5.26799\\
\hline
\multirow{2}{*}{60} & 0.001   & 0.11317 & 0.99129  & 0.991354 & 11.4026 & 11.624 \\
                    & 0.00375 & 0.40768 & 0.958162 & 0.95926  & 8.09264 & 8.75263 \\
                    & 0.0065  & 0.67689 & 0.914328 & 0.917377 & 6.08609 & 7.11274 \\
                    & 0.0085  & 0.85596 & 0.872908 & 0.87723  & 4.75373 & 5.97222 \\
                    & 0.0095  & 0.93987 & 0.8465   & 0.850894 & 4.01694 & 3.28954\\
\hline
\multirow{2}{*}{80} & 0.0005 & 0.07146 & 0.994861 & 0.994872 & 12.422  & 12.5517 \\
                    & 0.0015 & 0.21082 & 0.982067 & 0.982263 & 10.2319 & 10.6061 \\
                    & 0.0045 & 0.59998 & 0.932575 & 0.934483 & 7.21471 & 8.12184 \\
                    & 0.0065 & 0.83469 & 0.889489 & 0.893143 & 5.68847 & 6.91947 \\
                    & 0.0085 & 1.04846 & 0.831025 & 0.836144 & 3.97207 & 5.57342 \\
\end{tabular}
\end{ruledtabular}
\end{table}

\section{Conclusions}
\label{sec:conclusions}

In this work, we studied static dipolar boson stars, interpreted as bound states of two identical boson stars held in equilibrium by the balance between long-range gravitational attraction and short-range scalar repulsion. On the numerical side, we constructed sequences of isolated and dipolar boson stars in the solitonic model, with Noether charge $Q=40$, %$Q (=\frac{Q}{\mu^2\Phi_0^2})= 40$,
$60$, and $80$ (in units of $(\mu\Phi_0)^{2}$) for each individual star, by varying the gravitational coupling $G$ (in units of $\Phi_0^{-2}$) %$G (=G\,\Phi_0^2)$ 
for solutions along the relativistic branch. %We obtained the binding energy of the system by subtracting the mass of two isolated stars of charge $Q$ from that of the dipolar configuration of total charge $2Q$. \rg{I would remove this last sentence, in the end, we never show this "naive" binding energy in the paper, we just briefly mention it. What is important is the EOB effective energy $\gamma$ that is used for the comparison with the analytical model.}

On the analytical side, we described these configurations using the effective-one-body formalism.
{Starting from a recently derived first-post-Minkowskian accurate $w$-EOB scalar potential~\cite{Damour:2025oys}, we derived a corresponding energy-dependent Schwarzschild-like Lagrange-EOB metric potential $A(R,\gamma)= - g_{tt}^{\rm eff}(R,\gamma)$ by solving the algebraic-differential system defined by the two equations~\eqref{eq:rhoODE}, \eqref{eq:A1rho}. This gives a description of the interaction between two boson stars via a radial potential of the additive form \eqref{eq:Asplit} involving the sum of a purely gravitational (point-mass) contribution, $A^{\rm BH}{(R, \gamma)}$, and of a scalar contribution, $A^{\rm scalar}(R, \gamma)$.} 
While the gravitational contribution is given by the simple closed-form third-post-Minkowskian 
expression~\eqref{ABH}, the analytic expression of the Yukawa-like scalar contribution $A^{\rm scalar}(R, \gamma)$ defined by solving Eqs.~\eqref{eq:rhoODE}, \eqref{eq:A1rho} would be rather complicated. In practice we resorted to numerically solving the linear ODE~\eqref{eq:rhoODE}. %, \eqref{eq:A1rho}. \rg{Strictly speaking, only \eqref{eq:rhoODE} is an ODE. Replace by "In practice we resorted to numerically solving the linear ODE \eqref{eq:rhoODE}" ?} 
Figure \ref{fig:APotential} %displays two examples of 
illustrates the shape of the two
building blocks of our final $A(R,\gamma)$.
%derived the EOB potential in the $w$-EOB gauge for the scattering of two boson stars. However, static dipolar boson stars are bound configurations, and are hence more naturally described in a Schwarzschild-like gauge, similar to the traditional PN EOB. We therefore mapped the $w$-EOB potential onto the energy-dependent Schwarzschild-like LEOB gauge. Since both descriptions are based on a geodesic action, this map does not require a canonical transformation. Instead, it reduces to a simple energy-dependent radial-coordinate transformation between the isotropic coordinate of the $w$-EOB and the Schwarzschild-like coordinate of the LEOB. At first order in the scalar interactions, this gives a linear ODE, Eq.~\eqref{eq:rhoODE}, whose solution yields the scalar contribution to the LEOB potential.

In the Lagrange-EOB description, a static dipolar configuration then corresponds to an effective particle of mass $\mu_{\rm eff} = M_1 M_2/(M_1+M_2)$ at rest at a local minimum of the 
radial potential $A(R,\gamma)$. Since the potential depends on the effective energy $\gamma$, we determined the latter self-consistently by bootstrapping the equilibrium condition $\gamma^2 = \min_R A(R,\gamma)$, with the properties of the isolated boson star as input from our numerical solutions. For all three sequences, the analytical predictions agree quite well with the numerical results for the dipolar boson stars over the whole range of $G$. We also included the tidal contributions and found that their effect on $\gamma$ is negligible. {See  Figs.~\ref{fig:NRvsAR} (a--c) for a display of our results.}

As $G$ increases, the gravitational attraction grows relative to the scalar repulsion, and the minimum of the LEOB potential becomes shallower until it merges with a neighboring maximum into an inflection point. According to the analytical model, this critical configuration, defined by a triple zero of $\gamma^2 - A(R,\gamma)$, marks the end of the sequence of static equilibria corresponding to dipolar boson stars. The analytical prediction for this turning point lies close to the end of the numerical sequences, with a difference in $G$ ($|1-G_c^{\mathrm{analytical}}/G_c^{\mathrm{numerical}}|\times 100$) of $5.41$\% for $Q = 40$, $1.01$\% for $Q=60$ and $2.96$\% for $Q = 80$.  {See  Figs.~\ref{fig:NRvsAR} (a--c) where the end points of the analytical and numerical equilibrium curves are marked as dots.} %\rg{Not sure if it is necessary to refer again to Figs.~\ref{fig:NRvsAR} (a--c) here. We already mention that this figure summarizes our results in the previous paragraph.}
%\rg{We should put the difference values here, the numerical value of $G_c$ corresponds to the last line in the data files I gave you.} 

{The overall conclusion is that the Schwarzschild-like EOB description introduced here to describe the interaction of two boson stars accurately captures not only the binding energy of equilibrium dipolar boson stars, but also the maximal coupling beyond which no static configuration exists. %It would be interesting to explore 
A natural next step would be to explore whether it can also accurately capture the interaction between two boson stars in dynamical, gravitationally bound systems, such as inspiralling binaries.}

\section*{Data availability statement}
All numerical data underlying this work are available from the authors upon reasonable request.

\begin{acknowledgments}
T. J. is supported by a LabEx Junior Research Chair Fellowship at ENS, Paris. T.~J. and R. G. thank the hospitality and the
stimulating environment of the Institut des Hautes Etudes Scientifiques
(IHES).  The present research was partially supported by the 2021
Balzan Prize for Gravitation: Physical and Astrophysical Aspects,
awarded to T. Damour.  
\end{acknowledgments}

\bibliographystyle{hhieeetr}
\bibliography{prlref, biblio}

@Article{Friedberg1987,
  author    = {Friedberg, R. and Lee, T. D. and Pang, Y.},
  journal   = {Phys. Rev. D},
  title     = {Scalar soliton stars and black holes},
  year      = {1987},
  pages     = {3658--3677},
  volume    = {35},
  doi       = {10.1103/PhysRevD.35.3658},
  issue     = {12},
  numpages  = {0},
  publisher = {American Physical Society},
  url       = {https://link.aps.org/doi/10.1103/PhysRevD.35.3658},
}

@Article{Derrick1964,
  author    = {G. H. Derrick},
  journal   = {Journal of Mathematical Physics},
  title     = {{Comments on Nonlinear Wave Equations as Models for Elementary Particles}},
  year      = {1964},
  number    = {9},
  pages     = {1252--1254},
  volume    = {5},
  doi       = {10.1063/1.1704233},
  publisher = {{AIP} Publishing},
}

@Article{Ildefonso2023,
  author    = {P. Ildefonso and M. Zilhão and C. Herdeiro and E. Radu and N. M. Santos},
  journal   = {Physical Review D},
  title     = {Self-interacting dipolar boson stars and their dynamics},
  year      = {2023},
  issn      = {2470-0029},
  number    = {6},
  pages     = {064011},
  volume    = {108},
  doi       = {10.1103/physrevd.108.064011},
  publisher = {American Physical Society (APS)},
}

@Article{Tamaki2011,
  author    = {T. Tamaki and N. Sakai},
  journal   = {Physical Review D},
  title     = {{How does gravity save or kill Q-balls?}},
  year      = {2011},
  issn      = {1550-2368},
  number    = {4},
  pages     = {044027},
  volume    = {83},
  doi       = {10.1103/physrevd.83.044027},
  publisher = {American Physical Society (APS)},
}

@Article{Damour2009,
  author    = {Damour, Thibault and Nagar, Alessandro},
  journal   = {Physical Review D},
  title     = {Relativistic tidal properties of neutron stars},
  year      = {2009},
  issn      = {1550-2368},
  number    = {8},
  pages     = {084035},
  volume    = {80},
  doi       = {10.1103/physrevd.80.084035},
  publisher = {American Physical Society (APS)},
}

@Article{MR3043640,
  author   = {Hecht, F.},
  journal  = {J. Numer. Math.},
  title    = {New development in FreeFem++},
  year     = {2012},
  issn     = {1570-2820},
  number   = {3-4},
  pages    = {251--265},
  volume   = {20},
  fjournal = {Journal of Numerical Mathematics},
  mrclass  = {65Y15},
  mrnumber = {3043640},
}

@InBook{Maggio2021,
  author    = {Maggio, Elisa and Pani, Paolo and Raposo, Guilherme},
  pages     = {1--37},
  publisher = {Springer Singapore},
  title     = {{Testing the Nature of Dark Compact Objects with Gravitational Waves}},
  year      = {2021},
  isbn      = {9789811547027},
  booktitle = {Handbook of Gravitational Wave Astronomy},
  doi       = {10.1007/978-981-15-4702-7_29-1},
}

@Article{Gervalle2025,
  author    = {R. Gervalle and M. S. Volkov},
  journal   = {Physical Review D},
  title     = {Black holes with electroweak hair: The detailed derivation},
  year      = {2025},
  issn      = {2470-0029},
  number    = {6},
  volume    = {112},
  doi       = {10.1103/gzhg-62fz},
  publisher = {American Physical Society (APS)},
}

@article{BSFactory,
  author = {Romain Gervalle and
            Etevaldo dos Santos Costa Filho and
            V{\'i}ctor Jaramillo and
            Jorge F. M. Delgado and
            Carlos A. R. Herdeiro and
            Raimon Luna and
            Eugen Radu},
  title  = {{Boson Star Factory: Past, Present, and Future}},
  journal = {arXiv:2609.24913},
  eprint = "2609.24913",
  archivePrefix = "arXiv",
  primaryClass = "gr-qc",
  month = "9",
  year  = "2026"
}

@article{Cunha:2022tvk,
    author = "Cunha, P. and Herdeiro, C. and Radu, E. and Shnir, Ya.",
    title = "{Two boson stars in equilibrium}",
    eprint = "2210.01833",
    archivePrefix = "arXiv",
    primaryClass = "gr-qc",
    doi = "10.1103/PhysRevD.106.124039",
    journal = "Phys. Rev. D",
    volume = "106",
    number = "12",
    pages = "124039",
    year = "2022"
}

@article{Damour:2025uka,
    author = "Damour, Thibault and Nagar, Alessandro and Placidi, Andrea and Rettegno, Piero",
    title = "{Novel Lagrange-multiplier approach to the effective-one-body dynamics of binary systems in post-Minkowskian gravity}",
    eprint = "2503.05487",
    archivePrefix = "arXiv",
    primaryClass = "gr-qc",
    doi = "10.1103/41sd-g2gb",
    journal = "Phys. Rev. D",
    volume = "113",
    number = "2",
    pages = "024042",
    year = "2026"
}

@article{Damour:2025oys,
    author = "Damour, Thibault and Jain, Tamanna and Sperhake, Ulrich",
    title = "{Gravitational Scattering of Solitonic Boson Stars: Analytics vs Numerics}",
    eprint = "2512.00945",
    archivePrefix = "arXiv",
    primaryClass = "gr-qc",
    doi = "10.1103/24g1-k5p2",
    journal = "Phys. Rev. Lett.",
    volume = "137",
    number = "2",
    pages = "021401",
    year = "2026"
}

@Article{Evstafyeva2026,
  author    = {Evstafyeva, Tamara and Seifert, Antonia and Sperhake, Ulrich and Moore, Christopher J. and Jain, Tamanna},
  journal   = {Physical Review D},
  title     = {Lessons from binary dynamics of inspiralling equal-mass boson-star mergers},
  year      = {2026},
  issn      = {2470-0029},
  number    = {6},
  volume    = {114},
  doi       = {10.1103/k9qh-v6yy},
  publisher = {American Physical Society (APS)},
}

@Article{Marks2025,
  author    = {Marks, Gareth Arturo and Staelens, Seppe J. and Evstafyeva, Tamara and Sperhake, Ulrich},
  journal   = {Physical Review Letters},
  title     = {{Long-Term Stable Nonlinear Evolutions of Ultracompact Black-Hole Mimickers}},
  year      = {2025},
  issn      = {1079-7114},
  number    = {13},
  volume    = {135},
  doi       = {10.1103/lk48-7r2f},
  publisher = {American Physical Society (APS)},
}

@article{Schupp:1995dy,
    author = "Schupp, B. and van der Bij, J. J.",
    title = "{An axially symmetric Newtonian boson star}",
    eprint = "astro-ph/9508017",
    archivePrefix = "arXiv",
    reportNumber = "FREIBURG-THEP-95-18",
    doi = "10.1016/0370-2693(95)01327-X",
    journal = "Phys. Lett. B",
    volume = "366",
    pages = "85--88",
    year = "1996"
}

@Article{Kasuya2000,
  author    = {Kasuya, S. and Kawasaki, M.},
  journal   = {Phys. Rev. D},
  title     = {{Q-ball formation through the Affleck-Dine mechanism}},
  year      = {2000},
  issn      = {1089-4918},
  number    = {4},
  pages     = {041301},
  volume    = {61},
  doi       = {10.1103/physrevd.61.041301},
  publisher = {American Physical Society (APS)},
}

@Article{Lee1989,
  author    = {Lee, T. D. and Pang, Yang},
  journal   = {Nuclear Physics B},
  title     = {Stability of mini-boson stars},
  year      = {1989},
  issn      = {0550-3213},
  number    = {2},
  pages     = {477--516},
  volume    = {315},
  doi       = {10.1016/0550-3213(89)90365-9},
  publisher = {Elsevier BV},
}

@Article{Kusmartsev1991,
  author    = {Kusmartsev, Fjodor V. and Mielke, Eckehard W. and Schunck, Franz E.},
  journal   = {Physical Review D},
  title     = {Gravitational stability of boson stars},
  year      = {1991},
  issn      = {0556-2821},
  number    = {12},
  pages     = {3895--3901},
  volume    = {43},
  doi       = {10.1103/physrevd.43.3895},
  publisher = {American Physical Society (APS)},
}

@Article{Kusmartsev1992,
  author    = {Kusmartsev, Fjodor V. and Schunck, Franz E.},
  journal   = {Physica B: Condensed Matter},
  title     = {Analogies and differences between neutron and boson stars studied with catastrophe theory},
  year      = {1992},
  issn      = {0921-4526},
  number    = {1-4},
  pages     = {24--34},
  volume    = {178},
  doi       = {10.1016/0921-4526(92)90175-r},
  publisher = {Elsevier BV},
}

@Article{Vangstein:2026osl,
    author = "Vangstein, Selma and Moore, Christopher J.",
    title = "{On the Interpretation of Tests of the Horizon Area Law}",
    eprint = "2610.36162",
    archivePrefix = "arXiv",
    primaryClass = "gr-qc",
    month = "10",
    year = "2026"
}

@article{LIGOScientific:2025rid,
    author = "Abac, A. G. and others",
    collaboration = "LIGO Scientific, Virgo, KAGRA",
    title = "{GW250114: Testing Hawking{\textquoteright}s Area Law and the Kerr Nature of Black Holes}",
    eprint = "2509.08054",
    archivePrefix = "arXiv",
    primaryClass = "gr-qc",
    reportNumber = "LIGO-P2500421",
    doi = "10.1103/kw5g-d732",
    journal = "Phys. Rev. Lett.",
    volume = "135",
    number = "11",
    pages = "111403",
    year = "2025"
}

@article{Buonanno:1998gg,
  author = "Buonanno, A. and Damour, T.",
  title = "Effective one-body approach to general relativistic two-body dynamics",
  journal = "Phys. Rev. D",
  volume = "59",
  doi = "10.1103/PhysRevD.59.084006",
  note = "gr-qc/9811091",
  year = "1999",
  pages = "084006"}

@article{Buonanno:2000ef,
  author = "Buonanno, A. and Damour, T.",
  title = "{Transition from inspiral to plunge in binary black hole coalescences}",
  journal = "Phys. Rev. D",
  volume = "62",
  year = "2000",
  doi = "10.1103/PhysRevD.62.064015",
  pages = "064015",
  note = "gr-qc/0001013"}

@article{Bustillo:2020syj,
  author = "Bustillo, J. C. and Sanchis-Gual, N. and Torres-Forn\'e, A. and Font, J. A. and Vajpeyi, A. and Smith, R. and Herdeiro, C. and Radu, E. and Leong, S. H. W.",
  title = "{GW190521 as a Merger of Proca Stars: A Potential New Vector Boson of $8.7\times 10^{-13}$  eV}",
  eprint = "2009.05376",
  archivePrefix = "arXiv",
  primaryClass = "gr-qc",
  reportNumber = "LIGO DCC:P-2000353",
  doi = "10.1103/PhysRevLett.126.081101",
  journal = "Phys. Rev. Lett.",
  volume = "126",
  number = "8",
  pages = "081101",
  year = "2021"}

@article{CalderonBustillo:2022cja,
  author = "Calderon Bustillo, J. and Sanchis-Gual, N. and Leong, S. H. W. and Chandra, K. and Torres-Forne, A. and Font, J. A. and Herdeiro, C. and Radu, E. and Wong, I. C. F. and Li, T. G. F.",
  title = "{Searching for vector boson-star mergers within LIGO-Virgo intermediate-mass black-hole merger candidates}",
  eprint = "2206.02551",
  archivePrefix = "arXiv",
  primaryClass = "gr-qc",
  month = "6",
  year = "2022"}

@article{Cardoso:2017cfl,
    author = "Cardoso, Vitor and Franzin, Edgardo and Maselli, Andrea and Pani, Paolo and Raposo, Guilherme",
    title = "{Testing strong-field gravity with tidal Love numbers}",
    eprint = "1701.01116",
    archivePrefix = "arXiv",
    primaryClass = "gr-qc",
    doi = "10.1103/PhysRevD.95.084014",
    journal = "Phys. Rev. D",
    volume = "95",
    number = "8",
    pages = "084014",
    year = "2017",
    note = "[Addendum: Phys.Rev.D 95, 089901 (2017)]"
}

@article{Cardoso:2019rvt,
  author = "Cardoso, V. and Pani, P.",
  title = "{Testing the nature of dark compact objects: a status report}",
  journal = "Living Rev. Rel.",
  volume = "22",
  year = "2019",
  number = "1",
  pages = "4",
  doi = "10.1007/s41114-019-0020-4",
  eprint = "arXiv:1904.05363 [gr-qc]"}

@article{Carloni:2019cyo,
    author = "Carloni, Sante and Rosa, Jo\~ao Lu\'\i{}s",
    title = "{Derrick\textquoteright{}s theorem in curved spacetime}",
    eprint = "1906.00702",
    archivePrefix = "arXiv",
    primaryClass = "gr-qc",
    doi = "10.1103/PhysRevD.100.025014",
    journal = "Phys. Rev. D",
    volume = "100",
    number = "2",
    pages = "025014",
    year = "2019"
}

@article{Coleman:1985ki,
  author = "Coleman, S. R.",
  title = "{Q-balls}",
  reportNumber = "HUTP-85/A050",
  doi = "10.1016/0550-3213(86)90520-1",
  journal = "Nucl. Phys. B",
  volume = "262",
  number = "2",
  pages = "263",
  year = "1985",
  note = "[Addendum: Nucl.Phys.B 269, 744 (1986)]"}

@article{Cunha:2022gde,
  author = "Cunha, P. V. P. and Herdeiro, C. and Radu, E. and Sanchis-Gual, N.",
  title = "{Exotic Compact Objects and the Fate of the Light-Ring Instability}",
  eprint = "2207.13713",
  archivePrefix = "arXiv",
  primaryClass = "gr-qc",
  doi = "10.1103/PhysRevLett.130.061401",
  journal = "Phys. Rev. Lett.",
  volume = "130",
  number = "6",
  pages = "061401",
  year = "2023"}

@article{Damour:2022ybd,
    author = "Damour, Thibault and Rettegno, Piero",
    title = "{Strong-field scattering of two black holes: Numerical relativity meets post-Minkowskian gravity}",
    eprint = "2211.01399",
    archivePrefix = "arXiv",
    primaryClass = "gr-qc",
    doi = "10.1103/PhysRevD.107.064051",
    journal = "Phys. Rev. D",
    volume = "107",
    number = "6",
    pages = "064051",
    year = "2023"
}

@article{Evstafyeva:2024qvp,
  author = "Evstafyeva, T. and Sperhake, U. and Romero-Shaw, I. and Agathos, M.",
  title = "{Gravitational-wave data analysis with high-precision numerical relativity simulations of boson star mergers}",
  eprint = "2406.02715",
  archivePrefix = "arXiv",
  primaryClass = "gr-qc",
  doi = "10.1103/PhysRevLett.133.131401",
  journal = "Phys. Rev. Lett.",
  volume = "133",
  number = "13",
  pages = "131401",
  note = "arXiv:2406.02715 [gr-qc]",
  year = "2024"}

@article{Gleiser:1988rq,
  author = "Gleiser, Marcelo",
  title = "{Stability of Boson Stars}",
  reportNumber = "FERMILAB-PUB-88-067-A",
  doi = "10.1103/PhysRevD.38.2376",
  journal = "Phys. Rev. D",
  volume = "38",
  pages = "2376",
  year = "1988",
  note = "[Erratum: Phys.Rev.D 39, 1257 (1989)]"}

@article{Gleiser:1988ih,
  author = "Gleiser, Marcelo and Watkins, Richard",
  title = "{Gravitational Stability of Scalar Matter}",
  reportNumber = "NSF-ITP-88-152",
  doi = "10.1016/0550-3213(89)90627-5",
  journal = "Nucl. Phys. B",
  volume = "319",
  pages = "733--746",
  year = "1989"}

@article{Guzman:2009zz,
  author = "Guzman, F. S. and Rueda-Becerril, J. M.",
  title = "{Spherical boson stars as black hole mimickers}",
  journal = "Phys. Rev. D",
  volume = "80",
  year = "2009",
  pages = "084023",
  doi = "10.1103/PhysRevD.80.084023",
  eprint = "arXiv:1009.1250 [astro-ph]"}

@article{Kaup:1968zz,
  author = "Kaup, David J.",
  title = "{Klein-Gordon Geon}",
  journal = "Phys. Rev.",
  volume = "172",
  year = "1968",
  pages = "1331-1342",
  doi = "10.1103/PhysRev.172.1331"}

@article{Kleihaus:2005me,
  author = "Kleihaus, Burkhard and Kunz, Jutta and List, Meike",
  title = "{Rotating boson stars and Q-balls}",
  eprint = "gr-qc/0505143",
  archivePrefix = "arXiv",
  doi = "10.1103/PhysRevD.72.064002",
  journal = "Phys. Rev. D",
  volume = "72",
  pages = "064002",
  year = "2005"}

@article{Liebling:2012fv,
  author = "Liebling, S. L. and Palenzuela, C.",
  title = "{Dynamical boson stars}",
  eprint = "1202.5809",
  archivePrefix = "arXiv",
  primaryClass = "gr-qc",
  doi = "10.1007/s41114-023-00043-4",
  journal = "Living Rev. Rel.",
  volume = "26",
  number = "1",
  pages = "1",
  year = "2023"}

@article{Palenzuela:2006wp,
  author = "Palenzuela, C. and Olabarrieta, I. and Lehner, L. and Liebling, S. L.",
  title = "{Head-on collisions of boson stars}",
  journal = "Phys. Rev. D",
  volume = "75",
  year = "2007",
  pages = "064005",
  doi = "10.1103/PhysRevD.75.064005",
  eprint = "gr-qc/0612067"}

@article{Ruffini:1969qy,
  author = "Ruffini, Remo and Bonazzola, Silvano",
  title = "{Systems of selfgravitating particles in general relativity and the concept of an equation of state}",
  doi = "10.1103/PhysRev.187.1767",
  journal = "Phys. Rev.",
  volume = "187",
  pages = "1767--1783",
  year = "1969"}

@article{Schunck:1996he,
  author = "Schunck, F. E. and Mielke, E. W.",
  title = "{Rotating boson star as an effective mass torus in general relativity}",
  journal = "Phys. Lett.",
  volume = "A249",
  year = "1998",
  pages = "389-394",
  doi = "10.1016/S0375-9601(98)00778-6"}

@article{Schunck:2003kk,
  author = "Schunck, F. E. and Mielke, E. W.",
  title = "{General relativistic boson stars}",
  eprint = "0801.0307",
  archivePrefix = "arXiv",
  primaryClass = "astro-ph",
  doi = "10.1088/0264-9381/20/20/201",
  journal = "Class. Quant. Grav.",
  volume = "20",
  pages = "R301--R356",
  year = "2003"}

@article{Seidel:1993zk,
  author = "Seidel, E. and Suen, W.-M.",
  title = "{Formation of solitonic stars through gravitational cooling}",
  journal = "Phys. Rev. Lett.",
  volume = "72",
  pages = "2516-2519",
  doi = "10.1103/PhysRevLett.72.2516",
  year = "1994",
  note = "gr-qc/9309015"}

@article{Sennett:2017etc,
    author = "Sennett, Noah and Hinderer, Tanja and Steinhoff, Jan and Buonanno, Alessandra and Ossokine, Serguei",
    title = "{Distinguishing Boson Stars from Black Holes and Neutron Stars from Tidal Interactions in Inspiraling Binary Systems}",
    eprint = "1704.08651",
    archivePrefix = "arXiv",
    primaryClass = "gr-qc",
    doi = "10.1103/PhysRevD.96.024002",
    journal = "Phys. Rev. D",
    volume = "96",
    number = "2",
    pages = "024002",
    year = "2017"
}

@article{Siemonsen:2024snb,
  author = "Siemonsen, N.",
  title = "{Nonlinear Treatment of a Black Hole Mimicker Ringdown}",
  eprint = "2404.14536",
  archivePrefix = "arXiv",
  primaryClass = "gr-qc",
  doi = "10.1103/PhysRevLett.133.031401",
  journal = "Phys. Rev. Lett.",
  volume = "133",
  number = "3",
  pages = "031401",
  year = "2024"}

@book{Wald1984,
  author = "Wald, R. M.",
  title = "General {R}elativity",
  year = "1984",
  publisher = "The {U}niversity of {C}hicago {P}ress, {C}hicago and {L}ondon"}

@article{Yoshida:1997qf,
  author = "Yoshida, Shijun and Eriguchi, Yoshiharu",
  title = "{Rotating boson stars in general relativity}",
  reportNumber = "PRINT-97-136 (TOKYO)",
  doi = "10.1103/PhysRevD.56.762",
  journal = "Phys. Rev. D",
  volume = "56",
  pages = "762--771",
  year = "1997"}

@article{Yoshida:1997nd,
  author = "Yoshida, Shijun and Eriguchi, Yoshiharu",
  title = "{New static axisymmetric and nonvacuum solutions in general relativity: Equilibrium solutions of boson stars}",
  doi = "10.1103/PhysRevD.55.1994",
  journal = "Phys. Rev. D",
  volume = "55",
  pages = "1994--2001",
  year = "1997"}

@article{Abbott:2016blz,
  author = "Abbott, B. P. and others",
  title = "{Observation of Gravitational Waves from a Binary Black Hole Merger}",
  journal = "Phys. Rev. Lett.",
  volume = "116",
  year = "2016",
  number = "6",
  pages = "061102",
  doi = "10.1103/PhysRevLett.116.061102",
  eprint = "1602.03837",
  archivePrefix = "arXiv",
  primaryClass = "gr-qc",
  reportNumber = "LIGO-P150914"}

@article{Kim:2024wku,
    author = "Kim, Heejoo and Park, Junghyeon and Son, Minho",
    title = "{Axion dark matter from cosmic string network}",
    eprint = "2402.00741",
    archivePrefix = "arXiv",
    primaryClass = "hep-ph",
    doi = "10.1007/JHEP07(2024)150",
    journal = "JHEP",
    volume = "07",
    pages = "150",
    year = "2024"
}

@article{Inman:2019wvr,
  author = {Inman, D. and Ali-Ha{\"\i}moud, Y.},
  title = "{Early structure formation in primordial black hole cosmologies}",
  eprint = "1907.08129",
  archivePrefix = "arXiv",
  primaryClass = "astro-ph.CO",
  doi = "10.1103/PhysRevD.100.083528",
  journal = "Phys. Rev. D",
  volume = "100",
  number = "8",
  pages = "083528",
  year = "2019"}

@article{Vachaspati:1991tt,
    author = "Vachaspati, Tanmay and Vilenkin, Alexander",
    title = "{Large scale structure from wiggly cosmic strings}",
    reportNumber = "TUTP-91-5",
    doi = "10.1103/PhysRevLett.67.1057",
    journal = "Phys. Rev. Lett.",
    volume = "67",
    pages = "1057--1061",
    year = "1991"
}

@article{LIGOScientific:2026,
    author = "Abac, A. G. and others",
    collaboration = "LIGO Scientific, Virgo, KAGRA",
    title = "{GWTC-5.0: An Introduction to Version 5.0 of the Gravitational-Wave Transient Catalog}",
    eprint = "2605.27223",
    archivePrefix = "arXiv",
    primaryClass = "gr-qc",
    reportNumber = " LIGO-P2500701",
    month = "5",
    year = "2026"
}
\end{document}